\documentclass[aps,pre,twocolumn,superscriptaddress,twoside, nofootinbib]{revtex4-1}
\usepackage{amsmath,amssymb,amsfonts}
\usepackage{graphicx}
\usepackage[english]{babel}
\usepackage{psfrag}
\usepackage{color}
\usepackage{tabularx}
\usepackage{times}
\usepackage{hyperref}
\usepackage{float}

\begin{document}

\title{Tunable deconfined quantum criticality and interplay of different valence-bond solid phases}

\author{Bowen Zhao} 
\email{bwzhao@bu.edu} 
\affiliation{Department of Physics, Boston University, 590 Commonwealth Avenue, Boston, Massachusetts 02215, USA}

\author{Jun Takahashi} 
\email{jt@iphy.ac.cn}
\affiliation{Beijing National Laboratory for Condensed Matter Physics and Institute of Physics, Chinese Academy of Sciences, Beijing 100190, China}
\affiliation{Department of Physics, Boston University, 590 Commonwealth Avenue, Boston, Massachusetts 02215, USA}

\author{Anders W. Sandvik} 
\email{sandvik@bu.edu} 
\affiliation{Department of Physics, Boston University, 590 Commonwealth Avenue, Boston, Massachusetts 02215, USA}
\affiliation{Beijing National Laboratory for Condensed Matter Physics and Institute of Physics, Chinese Academy of Sciences, Beijing 100190, China}

\date{\today}

\begin{abstract}
We use quantum Monte Carlo simulations to study a quantum $S=1/2$ spin model with competing multi-spin interactions. We find a quantum phase 
transition between a columnar valence-bond solid (cVBS) and a N\'eel antiferromagnet (AFM), as in the scenario of deconfined quantum-critical 
points, as well as a transition between the AFM and a staggered valence-bond solid (sVBS). By continuously varying a parameter, the sVBS--AFM 
and AFM--cVBS boundaries merge into a direct sVBS--cVBS transition. Unlike previous models with putative deconfined AFM--cVBS transitions, e.g., 
the standard $J$-$Q$ model, in our extended $J$-$Q$ model with competing cVBS and sVBS inducing terms the transition can be tuned from continuous 
to first-order. We find the expected  emergent U(1) symmetry of the microscopically $Z_4$ symmetric cVBS order parameter when the transition is 
continuous. In contrast, when the transition changes to first-order the clock-like $Z_4$ fluctuations are absent and there is no emergent higher 
symmetry. We argue that the confined spinons in the sVBS phase are fracton-like. We also present results for an SU(3) symmetric model with a similar phase diagram. 
The new family of models can serve as a useful tool for further investigating open questions related to deconfined quantum criticality and its associated 
emergent symmetries.
\end{abstract}

\maketitle

\section{Introduction}
\label{intro}

The spin $S=1/2$ Heisenberg model with uniform nearest-neighbor interactions on the two-dimensional (2D) square lattice 
has a N\'eel antiferromagnetic (AFM) ordered ground state \cite{Manousakis91}. By introducing other interactions such as frustration or multi-spin 
interactions, the quantum fluctuation of the AFM order parameter can be increased, eventually destroying the AFM order and
leading to a different ground state. The long-wavelength behavior of Heisenberg and similar quantum magnets can be described field-theoretically by the 
O(3) nonlinear sigma model with a Berry phase term \cite{Haldane88,CHN89}. The Berry phase term will vanish for any smooth spin configurations, i.e., 
in the AFM phase, but will likely influence the phase diagram of the system if topological defects such as ``hedgehog" singularities are considered.  
In (2+1) dimension, such effects can play a dominant role and can drive the system into exotic paramagnetic phases, by which we mean collective 
many-body states with no direct classical analogues (see Ref.~\cite{Sachdev08} for a review). 

\subsection{Valence-bond solids}

\begin{figure}[b]
\includegraphics[width=65mm,clip]{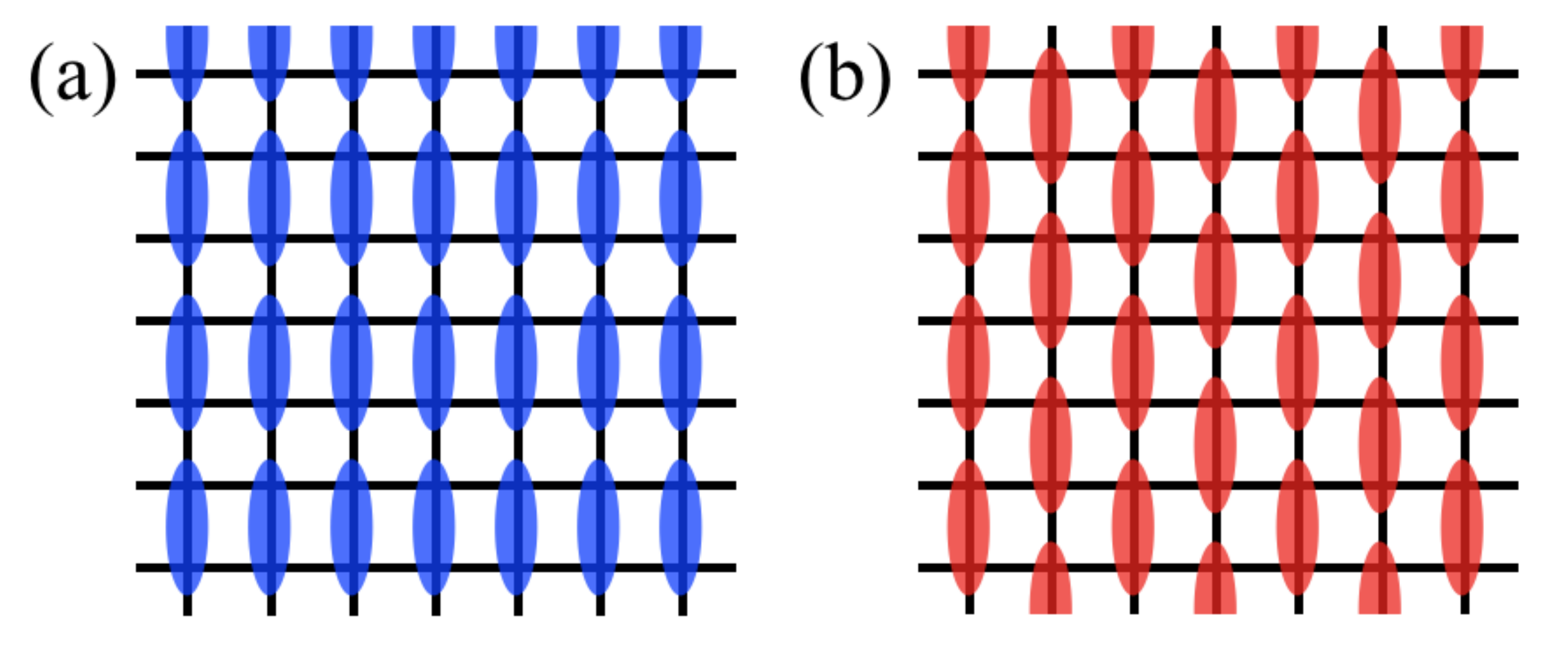}
\caption{Illustration of the dimer (singlet) patterns of a cVBS (a) and an sVBS (b). The symmetry properties of these patterns can be described 
by 2D space groups (wallpaper groups): the cVBS belongs to pmm, while the sVBS belongs to cmm.}
\label{Fig:Phase}
\end{figure}

One example of an interesting 2D paramagnetic phase is the so-called valence-bond solid (VBS) \cite{Read89,Read90,Dagotto89}, which preserves spin rotational 
symmetry but spontaneously breaks lattice symmetries through the condensation of singlets forming a regular pattern (or, more precisely, 
a modulation of the singlet density forms). There are several possible ways to break the lattice symmetries, and therefore VBS phases can also appear 
in many different incarnations. In this work we discuss two cases of dimer VBSs (i.e., the singlets form between two neighboring spins): the 
columnar VBS (cVBS) and staggered VBS (sVBS). As shown in Fig.~\ref{Fig:Phase}, in a cVBS phase the singlets align periodically with no ``shifting", 
thus breaking the symmetry group of the square lattice p4m down to pmm, while in the sVBS phase the singlets are successively shifted by one
lattice spacing  and form a stair-case pattern. This pattern belongs to the cmm space group. Though the symmetry groups are different, both these VBSs are 
four-fold degenerate on the infinite lattice or when placed on a torus with an even number of sites in both directions. We will discuss competition
between cVBS and sVBS ordering as well as the quantum phase transitions of these phases into the AFM state.
       
From the field theoretical perspective, the cVBS phase can be understood as a nontrivial magnetically disordered phase resulting from quantum 
tunneling of Skyrmions in a certain limit \cite{Read89,Read90,Murthy90}. One important consequence of this picture is the deconfined quantum 
criticality (DQC) scenario \cite{Senthil04}. By assuming the conservation of the Skyrmion number at the AFM--cVBS transition, it was proposed
that this transition can be described by a CP$^1$ field theory with non-compact U(1) gauge field. There are several interesting and controversial
assumptions and consequences of the DQC scenario, e.g., a generically continuous transition where normally a first-order transition would be expected (since the 
two ordered phases break unrelated symmetries), fractionalized excitations at the critical point, and a ``dangerously irrelevant'' $Z_4$ perturbing 
field with an associated emergent U(1) symmetry \cite{Levin04}. This type of order--order phase transition is beyond the conventional 
Landau-Ginzberg-Wilson (LGW) paradigm and has attracted intensive scrutiny during the past several years. 

\subsection{Deconfined quantum criticality in spin models}

Some elements of the DQC scenario being speculative, it is important to realize such phase transitions in concrete microscopic models, to fully test and 
explore the possible physics arising from the field theoretical proposal and beyond (i.e., features that were not part of the original DQC proposal). 
Since the sign-problem free $J$-$Q$ model \cite{Sandvik07} (with  Heisenberg exchange $J$ and a correlated multi-singlet projection $Q$) was proposed, many
detailed studies of the AFM--cVBS transition and the cVBS phase itself have been carried out with quantum Monte Carlo (QMC) simulations of this model 
\cite{Sandvik07,Melko08,Jiang08,Lou09,Sandvik10a,Kaul11,Sandvik12,Harada13,Chen13,Block13,Pujari15,Suwa16,Shao16,Ma18} and of classical 3D models exhibiting 
analogous transitions \cite{Kuklov08,Sreejith14,Nahum15a,Nahum15b}. These studies solved some previously open questions, but they also posed  new ones. For 
example observations of emergent higher symmetries at the critical point have been reported recently, with the AFM and cVBS order parameters combining
into a larger vector transforming under SO(5) \cite{Nahum15a,Suwa16} or O(4) \cite{Qin17,Ma19a} symmetry, depending on the model. The possibility of higher
symmetry was pointed out already some time ago \cite{Senthil06} and the numerical observations have further spurred the interest in this phenomenon, 
including in the context of the ``web of dualities'' between different field theories \cite{Wang17,Qin17}. Furthermore, a description of the transition
in terms of non-unitary complex conformal field theories (CFTs) \cite{Gorbenko18,Gorbenko18b,Ma19b,Nahum19} was inspired by unusual scaling behaviors observed in 
$J$-$Q$ and other models \cite{Jiang08,Kuklov08,Sandvik10a,Chen13,Sreejith14,Nahum15b}, which by some have been interpreted as signs of an eventually very 
weakly first-order transition in accessible models, with the critical point existing only in the complex plane. Alternatively, it has also been proposed that
the transition is continuous, as in the original DQC scenario, but with unusual scaling behavior stemming from two divergent length scales \cite{Shao16}.
It should be noted that there are no unambiguous signs of first-order transitions in the best candidate models. A further intriguing fact is that, in some
models where the AFM--VBS transition is clearly first-order, the coexistence state exhibits emergent O(4) \cite{Zhao19,Serna19} or SO(5) \cite{Takahashi20}
symmetry, instead of forming two distinct phases separated by tunneling barriers (in analogy with conventional classical co-existence states separated 
by free-energy barriers).

While the phenomenology of the DQC scenario does not rely on the AFM--VBS transition being strictly continuous (requiring in practice only that
the correlation length is very large, which has already been well established), it is still of fundamental interest to try to answer the following 
basic questions: (i) Are the candidate DQC transitions observed in simulations truly continuous, or do discontinuities and phase coexistence develop
on some large length scale? (ii) Is the emergent symmetry of the putative DQC point exact in the thermodynamic limit or does it break down at a finite 
length scale (i.e., even if the DQC transition itself is truly continuous)? (iii) Are the emergent O(4) and SO(5) symmetries observed in the coexistence 
states at some first-order AFM--VBS transitions asymptotically exact, or do they break down above some length scale? 

These questions are still hard to answer conclusively based on current numerical results, because they appear to involve exceedingly large length
scales (system sizes). In order to gain deeper understanding of the AFM--cVBS transition, it would likely be fruitful to construct models that enable 
tuning of the exotic properties and ingredients addressed in the theory of DQC points, and more broadly in scenarios of weakly first-order transitions with 
unusual properties. For example, suppressing the possible emergent symmetry of the transition is a promising approach, since it may give rise to a 
conventional strongly first-order transition that could be detectable with current numerical techniques. If the transitions can be tuned between continuous 
(or extremely weak first-order), to moderately weakly and strongly first-order, many of the intriguing phenomena listed above could be studied 
more systematically than what has been possible so far.

\begin{figure}[t]
\includegraphics[width=60mm,clip]{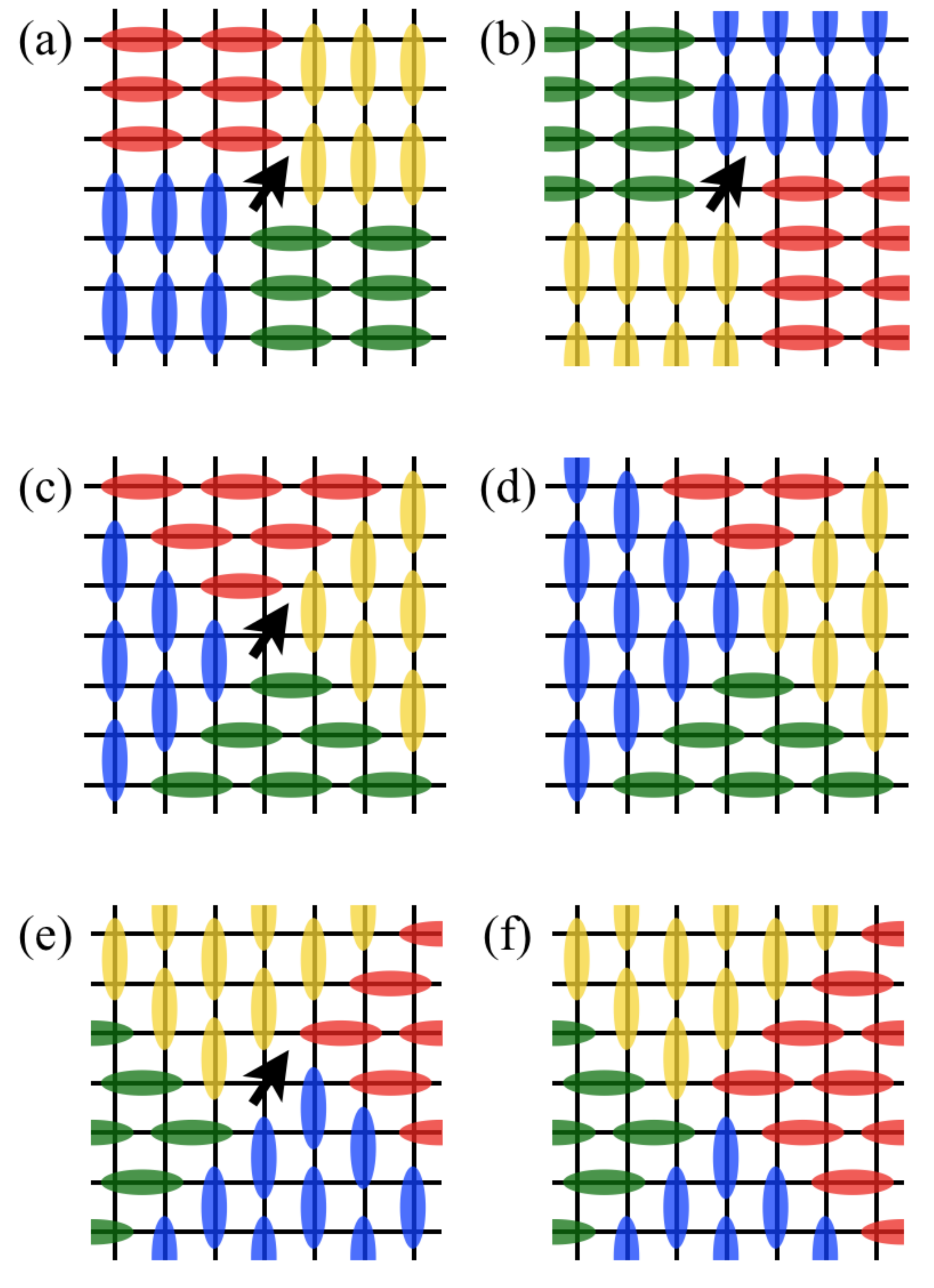}
\caption{Dimer patterns of different types of vortices in (a,b) the cVBS and in (c-f) the sVBS phase. 
In (a,b,c,e) the dimer patterns require an unpaired spin (spinon), 
while in (d,f) no spinon is induced.}
\label{Fig:Vertex}
\end{figure}

\subsection{Vortices and emergent symmetries}

The emergent U(1) symmetry of the DQC scenario [and also the possible higher spherical symmetry like SO(5)] is also related to the properties of the 
vortices in the VBS state. In the cVBS a nexus of four domain  walls separating the four different dimer patterns must necessarily have an unpaired 
spin \cite{Levin04}, i.e., the vortex is a spinon. Such spinons are bound into pairs in the cVBS phase and their deconfinement upon approaching the DQC 
point is directly related to the $Z_4$ vortices evolving into U(1) vortices. Even though the sVBS also is four-fold degenerate,  the $Z_4$ vortices 
in the sVBS can either carry a $S=1/2$ spinon or not, as illustrated in Fig.~\ref{Fig:Vertex}. Even if the spinon vortex is realized, unlike the case of
the cVBS the dimer pattern will only allow local fluctuations enabling the spinon to propagate in specific directions (in an extreme case of
only short dimers). The spinons may then be regarded as a type of fracton---an excitation with constrained mobility---as argued recently for spinons
in a plaquette VBS (pVBS) \cite{You20}. Though in the case of the pVBS the fracton property may only be realized in practice in extreme cases where
the four-spin singlets can be refarded as rigid objects, as pointed out in Ref.~ \cite{Takahashi20}, the phenomenon may actually be more easily 
realizable in an sVBS.

It would be interesting to investigate
whether competing sVBS interactions in a cVBS phase can suppress the fluctuation required for emergent U(1) symmetry and deconfined spinons, thus leading
to a different type of AFM--cVBS phase transition. We here desire to affect the properties of the vortices by tuning suitable microscopic couplings.
When the sVBS favoring interactions are sufficiently strong they may render the cVBS--AFM  transition first-order, unless there is a direct cVBS--sVBS
transition.  Conversely, the AFM--sVBS transition may possibly become less strongly first-order in the presence of the cVBS favoring interactions. 

Studying an sVBS phase and its transition to either an AFM phase or directly into a cVBS phase is interesting in its own right. 
There have been some discussions regarding phase transitions of the sVBS phase: Vishwanath et al.~proposed a possible continuous phase transition 
in bilayer honeycomb lattice \cite{Vishwanath04},~and Xu et al.~studied certain types of transitions between the sVBS and a $Z_2$ spin Liquid or 
an AFM phase \cite{Xu11}. Numerically, an sVBS state was studied with a $J$-$Q_3$ model where the three singlet projectors in the $Q_3$ terms are 
arranged in a stair-case fashion for the square lattice \cite{Sen10}. 
%and with a $J$-$Q_2$ model where the $Q_2$ terms are projectors of two parallel singlets for the honeycomb lattice. 
The AFM--sVBS transition in this case as well as in other lattices \cite{Banerjee11} are very strongly first-order, and the sVBS phase 
itself exhibits only weak fluctuations from the maximally ordered dimer singlet configuration. 
Therefore, the model in its current form does not offer many 
insights into the more interesting case of a strongly fluctuating sVBS where topological defects can play a more prominent role. 
Considering the 
possibility of fracton-type spinons and their propensity to nucleate AFM order (in analogy with Ref.~\cite{You20}), it is, however, not even clear
if very strong quantum fluctuations of the sVBS can be achieved before a first-order transition takes place. 

\subsection{Desired model properties}

Since we are interested in the generic properties of the VBS phases, we will take the approach of ``designer Hamiltonians" 
\cite{Kaul13}. The central idea here is to construct an easy-to-study model (e.g., without sign problems in QMC simulations) and to 
analyze particular phases and phase transitions. The concepts of universality and renormalizaion-group fixed points ensure that the low-energy properties 
of interest are still of relevance to real-world applications, even though interactions that can be realized in real materials are not faithfully 
represented microscopically. The following is a list of desired features for the designer Hamiltonian in our study: 
	
	\begin{itemize}
		\item harbors all of cVBS, sVBS, and AFM phases;
		\item the cVBS--AFM transition can be tuned from continuous or weakly-first order to more strongly first order;
		\item realizes a direct cVBS--sVBS transition as well as cVBS and sVBS in parts of its phase diagram.  
	\end{itemize}
	
In previous studies, none of the existing models were able to connect the first-order sVBS--AFM transition and the cVBS--AFM DQC transition, nor 
could they realize a direct transition between the two VBS states. In this work, we introduce a model Hamiltonian with two competing terms that individually 
favor a cVBS and an sVBS ground state, respectively, as discussed above. The motivation here is that the presence of sVBS favoring interactions in the cVBS phase
may to some degree suppress the development of the emergent U(1) symmetry that characterizes the DQC transition between the cVBS and the AFM.
We will show here that these expectations are borne out by our results for the model we have constructed, in particular as regards the evolution 
of the AFM--cVBS transition from continuous or very weakly first-order to clearly first-order.

\subsection{Article outline}

	The structure of the remainder of this article is as follows: 
	In Sec.~\ref{Sec:Model} we introduce the $S=1/2$ [SU(2)] models we study in this work. 
	In Sec.~\ref{Sec:Properites} we extract the phase diagram of the model and discuss the nature of the various quantum phase transitions we have identified.
	In Sec.~\ref{Sec:SU3Model} we discuss the SU(N) generalization of our model and present results for $N=3$.
	In Sec.~\ref{Sec:Discussion} we summarize our results and discuss some possible further applications of the models.

\section{Model and observables}

	\label{Sec:Model}
	\begin{figure}[t]
		\includegraphics[width=65mm,clip]{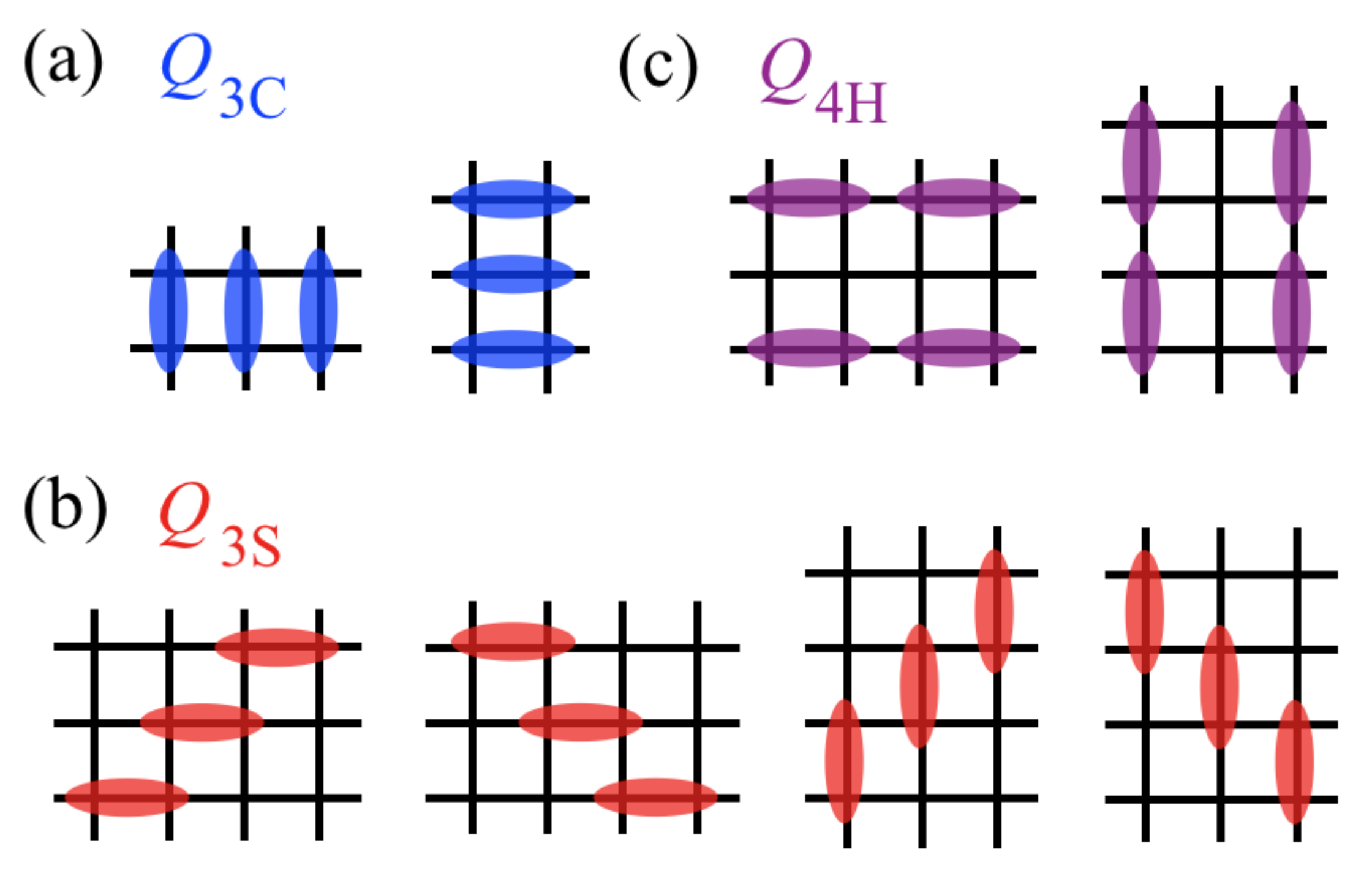}
		\caption{Depictions of the interactions included in the quantum spin model studied in this work. 
                %Black and white circles represent the two sub-lattices of the square lattice. 
		  The nearest-neighbor Heisenberg exchange ($J$) interactions act on all bonds on the square lattice drawn in black here. 
		  The different arrangements of the $n$ singlet projectors ($n=3,4$) in the multi-projector product terms are 
		  columnar $n=3$ of strength $Q_{3{\mathrm C}}$ in (a), staggered $n=3$ of strength $Q_{3{\mathrm S}}$ in (b), 
		  and $n=4$ with a larger separation of the projectors in (c) with strength $Q_{4{\mathrm H}}$.
                The Hamiltonian contains all spacial translations of the projectors so that the full square-lattice symmetry is maintained. }
		\label{Fig:models}
	\end{figure}

	We start from two previously studied designer Hamiltonians that host cVBS and sVBS phases, respectively. 
	The ground state of a columnar $J$-$Q_{3{\mathrm C}}$ model \cite{Lou09} goes through a DQC type cVBS--AFM transition, 
	and the ground state of a staggered $J$-$Q_{3{\mathrm S}}$ model \cite{Sen10} exhibits a first-order sVBS--AFM transition.
        Here and henceforth we use the second subscript to indicate a columnar (C) or staggered (S) spatial arrangement of the 
        three singlet projectors used in the multi-spin interaction. A natural strategy to construct a model that harbors both cVBS and 
        sVBS ground states would be to combine the $Q_{3{\mathrm C}}$ terms and $Q_{3{\mathrm S}}$ terms as follows: 
	\begin{eqnarray}
	\label{Eq:Model_Q3Q3s}
		H =
		 - Q_{3{\mathrm C}}\sum_{\mathrm C} P_{ij} P_{kl} P_{mn} - Q_{3{\mathrm S}}\sum_{\mathrm S} P_{ij} P_{kl} P_{mn},
	\end{eqnarray}
	where $P_{ij}$ denote a projection operator to a spin singlet state on sites $i$ and $j$, $P_{ij}=1/4-{\bf S}_i \cdot {\bf S}_j$,
        and the summation indices $\mathrm{C}$ and $\mathrm{S}$ stand for all arrangements of the six indices $i$-$m$ in columns or stairs, respectively.
        For details of the placement of the indices we refer to Figs.~\ref{Fig:models}(a) and (b). 
	
	In the two extreme cases where $Q_{3{\mathrm C}}/Q_{3{\mathrm S}}\rightarrow \infty$ and $Q_{3{\mathrm S}}/Q_{3{\mathrm C}}\rightarrow \infty$
        in Eq.~(\ref{Eq:Model_Q3Q3s}) 
	the ground states of $H$ would be in the cVBS and sVBS phase, respectively. However, there is not necessarily a direct transition
        between these two VBS phases with just the terms in the Hamiltonian defined in Eq.~(\ref{Eq:Model_Q3Q3s}). As we will see below, there 
        is an AFM phase between them, i.e., AFM order can form as a compromise between two competing VBS states.

        In order to achieve a direct cVBS--sVBS transition, we have to introduce another interaction. The Heisenberg exchange $J$ only strengthens
        the AFM order and expands the AFM phase between the two VBS phases. We will therefore not further discuss the Heisenberg interaction here
        and instead consider another type of multi-spin $Q$-type interaction. From the field theoretical point of view, we would like to turn on some 
        symmetry-allowed relevant terms with respect to the cVBS--AFM fixed point. An interaction achieving our aim is the $Q_{4{\mathrm H}}$ terms with the 
        relative arrangement of the four singlet projectors illustrated in Fig.~\ref{Fig:models}(c), which has some resemblance to the letter H
        that we use to identify this interaction. By combining it with the two previous terms we obtain
	\begin{eqnarray}
	\label{Eq:Model_Q4}
	H &=& - Q_{3{\mathrm C}}\sum_{{\mathrm C}} P_{ij} P_{kl} P_{mn} - Q_{3{\mathrm S}}\sum_{\mathrm S} P_{ij} P_{kl} P_{mn} \nonumber\\
	&& - Q_{4{\mathrm H}}\sum_{\mathrm H} P_{ij} P_{kl} P_{mn} P_{pq},
	\end{eqnarray}
        which is the model we study in this paper.

	Since the $Q_{4{\mathrm H}}$ terms are compatible with both cVBS and sVBS order (i.e., a subset of the terms have all their projectors on strong bonds in the
        respective phases), we expect that they should strengthen both the cVBS and sVBS 
        orders, thereby suppressing the AFM phase. This is indeed the case, and, moreover, we will find that $Q_{4{\mathrm H}}$ in combination
        with $Q_{3{\mathrm S}}$ can change the nature of the cVBS--AFM transition from a DQC transition to a clearly first-order transition.

       To study all three phases we define their order parameters as follows:
        \begin{subequations}
	\begin{eqnarray}
		m_z &=& \frac{1}{L^2} \sum_{\mathbf{r}} (-1)^{\mathbf{r}_{x}+\mathbf{r}_{y}} S^z (\mathbf{r}),\label{mzdef} \\
		m_c &=& \frac{1}{L^2} \sum_{\mathbf{r}}  (-1)^{\mathbf{r}_{x}} P^z_{\mathbf{r}, \mathbf{r}+\hat{x}} + 
                   i  (-1)^{\mathbf{r}_{y}} P^z_{\mathbf{r}, \mathbf{r}+\hat{y}}, \label{mcdef} \\
		m_s &=& \frac{1}{L^2} \sum_{\mathbf{r}}   (-1)^{\mathbf{r}_{x} + \mathbf{r}_{y}} P^z_{\mathbf{r}, \mathbf{r}+\hat{x}} + 
              i  (-1)^{\mathbf{r}_{x} + \mathbf{r}_{y}} P^z_{\mathbf{r}, \mathbf{r}+\hat{y}}, \label{msdef}~~~~ 
	\end{eqnarray}
        \label{m2defs}
        \end{subequations}
	where $\mathbf{r}$ runs over all sites of the $L \times L$ square lattice with periodic boundary conditions and $P_z$ is the $z$-component of 
        the singlet projector; $P^z_{{\bf r}_1,{\bf r}_2}=1/4-S^z_{{\bf r}_1}S^z_{{\bf r}_2}$, where for convenience we now use the lattice coordinates
        ${\bf r}$ as indices instead of the site indices used in the  Hamiltonian, Eq.~(\ref{Eq:Model_Q4}).
        We can further define the Binder cumulants to capture the characteristic order-parameter distributions in the different phases;
        \begin{subequations}
	\begin{eqnarray}
		U_z &=& \frac{5}{2} \left(1 - \frac{\langle m_z^4 \rangle}{3\langle m_z^2 \rangle^2} \right), \label{uzdef} \\
		U_{c,s} &=& 2 \left(1 - \frac{\langle m_{c,s}^4 \rangle}{2\langle m_{c,s}^2 \rangle^2} \right),
	\end{eqnarray}
        \label{udefs}
        \end{subequations}
        where the coefficients are chosen according to the number of components of the order parameter (and in the case of $U_z$ also taking
        into account that only one out of three components of the staggered magnetization is used), such that the cumulants approach $1$ with
        increasing $L$ if there is order of the given type and $0$ else. The condition $U_i \to 1$ is only relying on the presence of long range
        order, while $U_i \to 0$ in a disordered phase also relies on the fluctuations being Gaussian. As we will see below, there are cases where the
        fluctuations are not Gaussian even though the correlations are short-ranged.
	
        We used the Stochastic Series Expansion (SSE) QMC method for all the calculations presented here. For details of this algorithm         
        we refer to the review Ref.~\onlinecite{Sandvik10b}, which includes an implementation for the conventional $J$-$Q$ model. The more complicated
        $Q$ terms used here can be treated with simple generalizations.	
	
\section{Numerical Results}
\label{Sec:Properites}

\begin{subequations}
To investigate the ground state phase diagram of the model, Eq.~(\ref{Eq:Model_Q4}), with its three parameters $Q_{3{\mathrm C}}$, $Q_{3{\mathrm S}}$, 
and $Q_{4{\mathrm H}}$, we consider the plane of two dimensionless coupling ratios $(g,h)$:
\begin{eqnarray}
g&=&Q_{3{\mathrm C}} / (Q_{3{\mathrm C}} + Q_{3{\mathrm S}}),\label{gdef}\\
h&=&Q_{4{\mathrm H}} /(Q_{3{\mathrm C}} + Q_{3{\mathrm S}} + Q_{4{\mathrm H}}). \label{hdef}
\end{eqnarray}
To fix the energy scale, we always keep $Q_{3{\mathrm C}} + Q_{3{\mathrm S}} + Q_{4{\mathrm H}}=1$, and in the simulations with this convention we set the inverse
temperature to $\beta=L$.
\end{subequations}

\begin{figure*}[t]
\includegraphics[width=178mm]{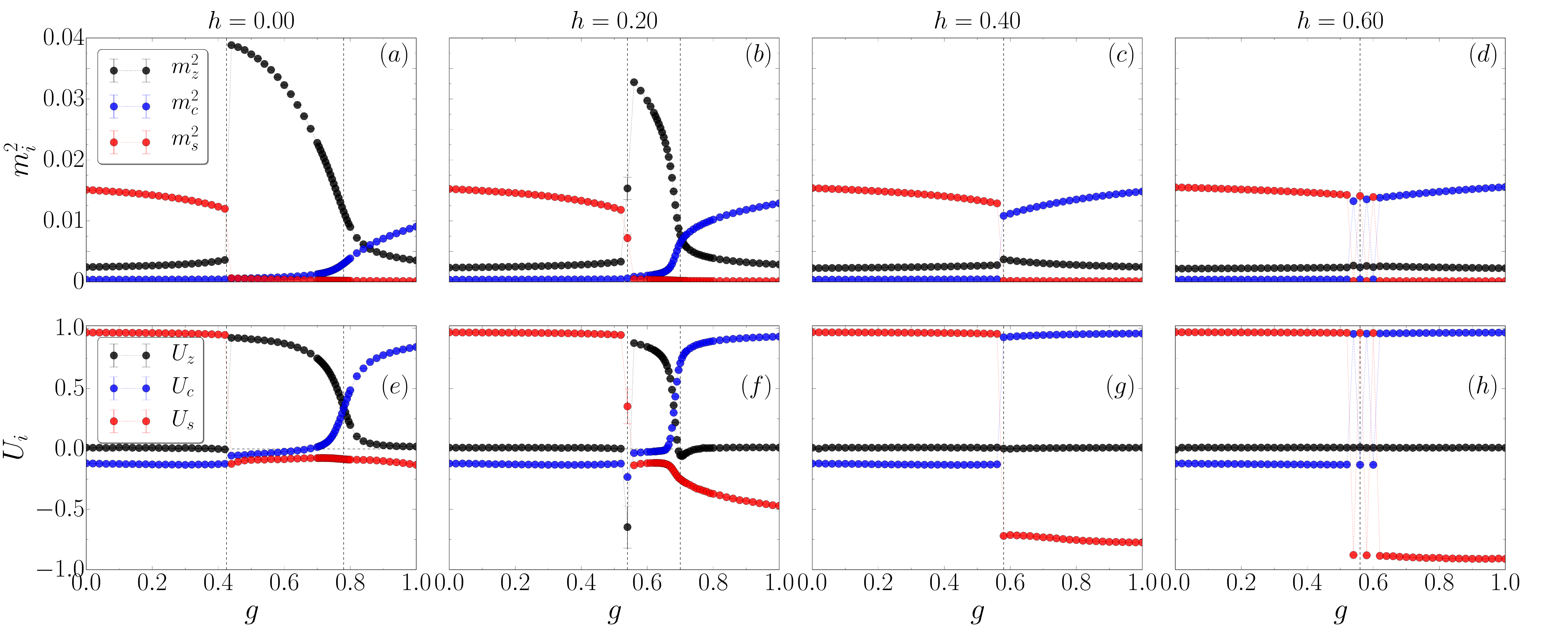}
\caption{The order parameters defined in Eqs.~(\ref{m2defs}) and their corresponding Binder ratios, Eqs.~(\ref{udefs}).
The results were obtained by SSE calculations of $L=16$ systems at inverse temperature $\beta=L$. From left to right panels, the
coupling ratio defined in Eq.~(\ref{hdef}) is $h = 0, 0.2, 0.4, 0.6$. We use the same colors as in Fig. \ref{Fig:PhaseDiagram} 
to indicate the different phases. Large fluctuations close to the phase transition at $h=0.6$ are due to simulations trapped
in meta-stable states, as is typical in simulations of strongly first-order transitions.}
\label{Fig:Q4OrderPara}
\end{figure*}

\begin{figure}[b]
\includegraphics[width=75mm,clip]{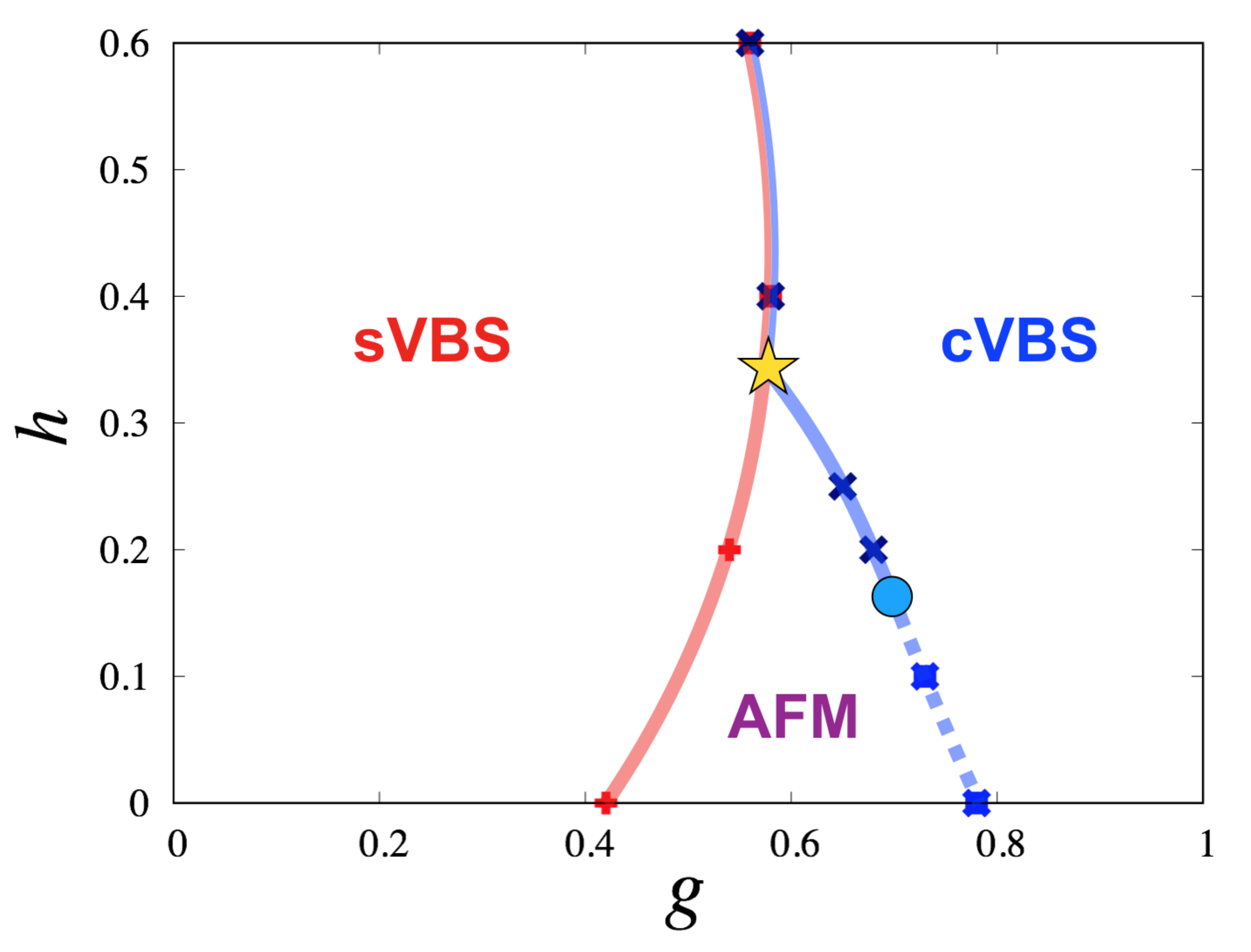}
\caption{Ground state phase diagram of the Hamiltonian Eq.~(\ref{Eq:Model_Q4})
in the plane of coupling ratios defined in Eqs.~(\ref{gdef}) and (\ref{hdef}). The phase boundaries are 
based on scans vs $g$ for for several values of $h$, as exemplified by Figs.~\ref{Fig:Q4OrderPara}, \ref{Fig:DetailedOrderPara_0}, 
and \ref{Fig:DetailedOrderPara_1}. The resulting transition points are indicated with $+$ and $\times$ symbols and the curves are smooth interpolations
and extrapolations of these points. The dashed and solid curves denote first-order transitions, while the solid curve segment denotes a
continuous (or possibly very weakly first-order) DQC transition. The three first-order phase boundaries meet at a triple point denoted 
by the yellow star, and there is possibly a tricritical point on the AFM--cVBS curve, with location approximately at the blue circle.}
\label{Fig:PhaseDiagram}
\end{figure}

\subsection{Phase diagram}

In Fig.~\ref{Fig:Q4OrderPara} we present results for the order parameters and Binder cumulants versus $g$ for several values
of $h$. Here we use a rather small system, $L=16$, but this already gives us an initial impression about the phase diagram.
Fig.~\ref{Fig:PhaseDiagram} shows a phase diagram based on several extracted transition points, using also results for systems 
larger than $L=16$ close to the transition point (some of which are further discussed below), for extrapolating to the thermodynamic 
limit.

We observe the three different phases as expected. A key feature here is that the $Q_{4{\mathrm H}}$ interaction ($h>0$) indeed shrinks the
AFM phase and eventually brings the two VBS phases into contact with each other, thus supporting the intuition that the specific
arrangement of the projectors in Fig.~\ref{Fig:models}(c) favor both types of VBS phases while suppressing the AFM phase.  The direct 
phase transition between the two VBS phases starts above some value of $h$ close to $0.35$ is strongly first-order, as also would be 
expected on the grounds that the two phases break the lattice symmetry in different ways. 

The sVBS--AFM transition is always first-order, as is clear from the sharp jumps in both the order parameters and Binder
cumulants in Fig.~\ref{Fig:Q4OrderPara} for $h=0.4$ and $0.6$. The behavior is similar to the previously studied $J$-$Q_{3{\mathrm S}}$
model \cite{Sen10}. The first-order transition is expected since, as discussed in Sec.~\ref{intro}, the DQC scenario does not apply 
to the sVBS state. We only observe a mild reduction of the size of the first-order discontinuities with increasing $h$. The location of
the triple point where all the phases come together still has some uncertainty, due to difficulties with the size extrapolations close 
to this point.
	
\subsection{Nature of the AFM--cVBS transition}
\label{sec:afmcvbs}

\begin{figure*}[t]
\includegraphics[width=178mm]{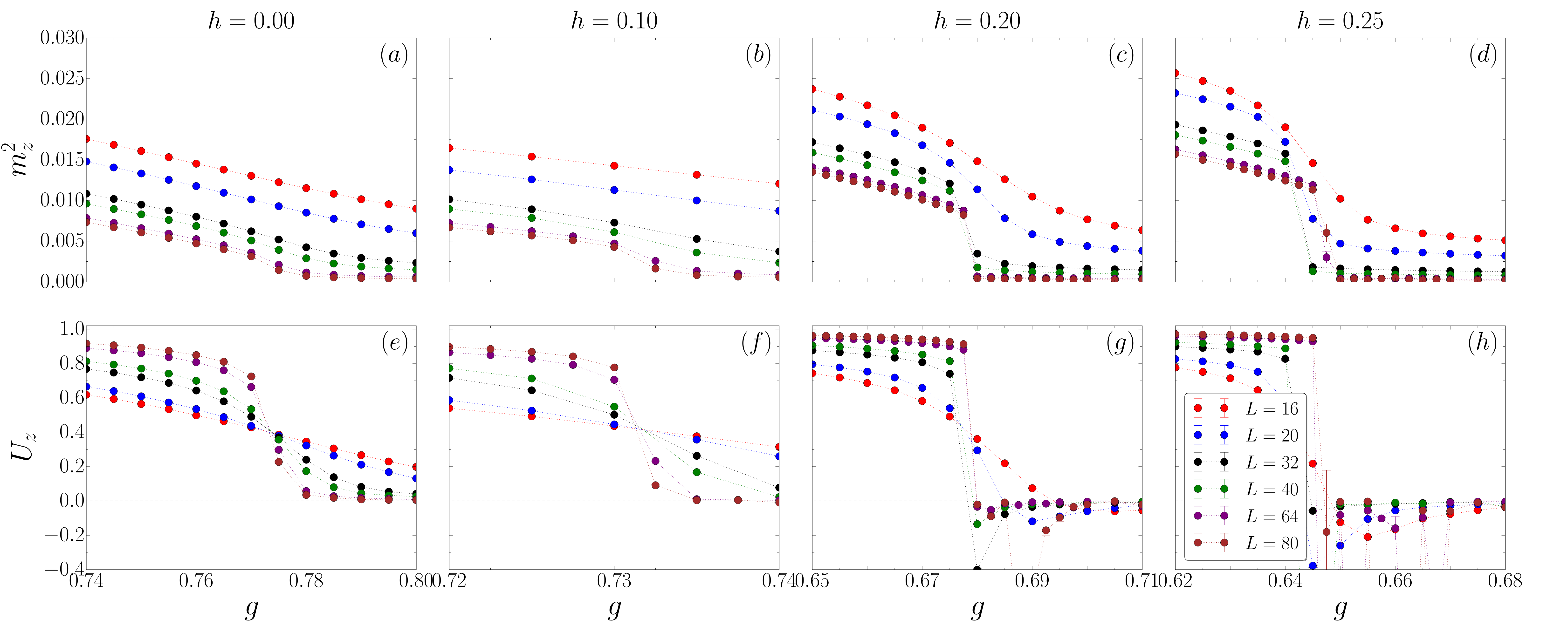}
\caption{Results near the cVBS--AFM transition for $h=0$, $0.1$, $0.2$, and $0.25$, from left to right as indicated.
Squared AFM order parameter are shown in panels (a)-(d) and the  corresponding Binder cumulant in (e)-(h). Results for different
system sizes are color coded as indicated in the legends in (h). In the cVBS phase for 
$h=0.25$ the simulations again some times get trapped in states with defects, as in Fig.~\ref{Fig:Q4OrderPara}, which causes sharp 
deviations from the correct ground state cumulant $U_z$.}
\label{Fig:DetailedOrderPara_0}
\end{figure*}

\begin{figure*}[t]
\includegraphics[width=178mm]{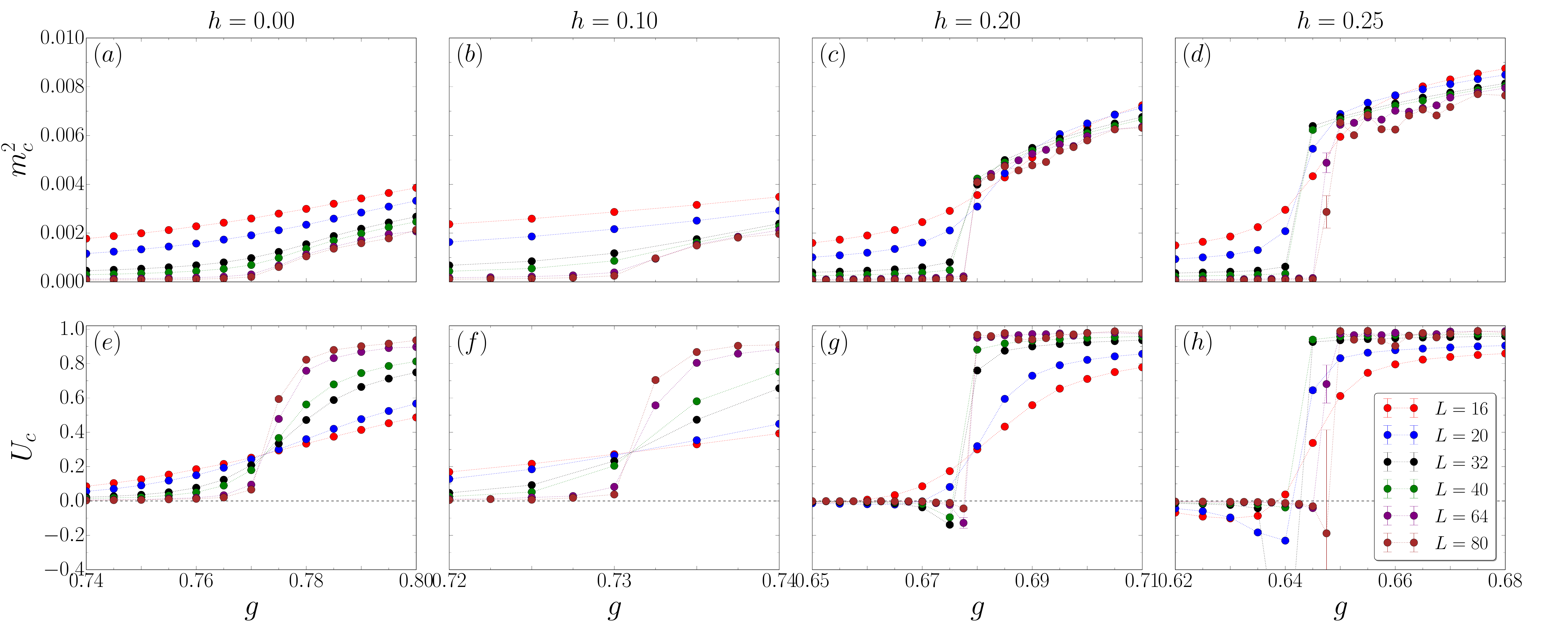}
\caption{Results for the squared cVBS order parameter in (a)-(d) and the corresponding Binder cumulants in (e)-(h),
obtained in the same simulations as in Fig.~\ref{Fig:DetailedOrderPara_0}.}
\label{Fig:DetailedOrderPara_1}
\end{figure*}
	
As discussed in Sec.~\ref{intro}, the AFM--cVBS transition in the standard $J$-$Q$ models (the $J$-$Q_{2C}$ and $J$-$Q_{3{\mathrm C}}$ models in the
notation used here) are of the DQC type, possibly with extremely weak and currently not detectable discontinuities. The extended model used 
here does not reduce to the previously studied cases because of the lack of $J$ terms and also because we keep $Q_{3{\mathrm S}}+Q_{3{\mathrm S}} + Q_{4{\mathrm H}}=1$,
so that there is always some $Q_{3{\mathrm S}}$ and (for $h>0$) $Q_{4{\mathrm H}}$
interaction present at the AFM--cVBS transition. Nevertheless, based on the $L=16$ results shown in Fig.~\ref{Fig:Q4OrderPara}, the 
transition looks continuous for $h=0$ and $0.2$ (where there is still an AFM phase). However, in order to determine the true nature of the transition 
the system-size dependence has to be studied very carefully and on much larger lattices. Here our aim is not to draw definite conclusions, as even 
in the standard $J$-$Q$ models the situation is not fully settled and very significant computational resources are required in order to obtain 
size-converged critical exponents or, alternatively, to detect weak discontinuities. 

As an initial study of the AFM--cVBS transition in the extended $J$-$Q$ model, in Figs.~\ref{Fig:DetailedOrderPara_0} and \ref{Fig:DetailedOrderPara_1} 
we present results for AFM and cVBS quantities, respectively. We focus on relatively small $h$ values, $h=0,0.1,0.2$, and $0.25$, for which the transition 
is either continuous or weakly first order, and in each case we show results for several different system sizes between $L=16$ and $L=80$. 
For $h=0$ and $0.1$, we find that both order parameters evolve smoothly versus $g$ for all system sizes, while for the larger $h$ values 
obvious discontinuities develop with increasing $L$. For the smaller $h$ values the Binder cumulants for different $L$ cross each other in 
what appears to be a single point (though one should expect some drifts in the crossing points if the behaviors are examined in more detail
with high-quality data on a denser grid), which is characteristic for continuous transitions. For the larger $h$ values the cumulants 
are more sharply varying and also become negative in the neighborhood of the transition. The latter behavior suggests that the cumulants will
eventually, for larger system sizes, develop the sharp negative peaks that are characteristic for conventional first-order transitions. Such
peaks should asymptotically diverge in proportion to the system volume, as previously observed with the $J$-$Q_{3{\mathrm S}}$ model \cite{Sen10}.

Recently, unconventional first-order transitions with emergent spherical symmetry of the order parameters were identified where no cumulant
minimums are observed, which can be traced to the lack of tunneling barriers when the two order parameters can be continuously rotated into
each other \cite{Zhao19,Serna19,Takahashi20}. The results for $h\ge 0.2$ in Figs.~\ref{Fig:DetailedOrderPara_0} and \ref{Fig:DetailedOrderPara_1}
suggest that the transitions here are not of the kind with emergent symmetry (though we have not established the volume-proportionality of the
cumulant peak growth). In Sec.~\ref{sec:vbsfluct} we will also explicitly confirm that there is no emergent symmetry of the two order parameters.
For $h=0$ and $0.1$, we can only say with certainty that the transitions are less first-order-like than at the higher $h$ values, as it is possible 
that discontinuities and negative cumulant peaks develop above some system size that decreases with increasing $h$. 

Note again that, also
at $h=0$ there are still $Q_{3{\mathrm S}}$ interactions present in the $(g,h)$ parameterization of the model that we have used here. It is possible,
in principle, that the staggered term is relevant and alters the DQC transition immediately, though we see no evidence of such behavior here
(and the prospect also appears unlikely, as there is no apparent new scaling field that this operator can bring in that is originally absent).
To compute the scaling dimension of the $Q_{3{\mathrm S}}$ coupling is possible in principle by analyzing the corresponding correlation function at
the AFM-cVBS transition of the standard $J$-$Q$ model. However, with the large number of singlet projectors involved (six in total for the 
correlation of the $Q_{\rm S3}$ terms) involved such a calculation would be very challenging \cite{Beach06} and we have not attempted this.

\subsection{Fluctuations in the VBS phases}
\label{sec:vbsfluct}

Consider the global cVBS angle $\phi$ defined from the complex order parameter $m_c$ in Eq.~(\ref{mcdef}) written as $m_c=|m_c|{\rm e}^{i\phi}$.
Starting from one out of the four dimer patterns in the cVBS phase, a series of local dimer flips (i.e., two adjacent parallel dimer rotated
spatially by $\pi/2$) can gradually shift the angle. Similarly, by replacing a pair of parallel dimers by a superposition of $x$- and $y$-oriented
dimers (i.e., plaquette singlets),
the angle also shifts relative to that with only static columnar dimers. Since the dimer order in the cVBS phase undergoes significant
fluctuations (not only of the kind affecting short dimers as discussed above, but the actual state also contains longer bipartite valence bonds), 
neither the angle $\phi$ nor the magnitude $|m_c|$ is fixed, but their values as obtained from SSE configurations fluctuate. The relative size 
of the fluctuations vanish as $L \to \infty$ as long as the system is cVBS ordered in the thermodynamic limit.

The order parameter can also be defined on some cell of size $\lambda$ smaller than the lattice size $L$, and, in the same way as explained above, 
even in an cVBS state on an infinite lattice the local angle $\phi({\bf r})$ thus defined in a cell centered at ${\bf r}$ can take values also between 
the four columnar angles $n\pi/2$, $n=0,1,2,3$. When crossing a $\pi/2$ domain wall, $\phi({\bf r})$ rotates between two adjacent values of $n$ over the 
width of the domain
wall, and such intermediate angles correspond to the presence of resonating valence bonds. In the center of a $\pi/2$ domain wall between two columnar 
states the dimer patterns is exactly that of a pVBS state, with resonating pairs of horizontal and vertical dimers forming on a specific 
plaquette pattern (one out of four possible patterns, depending on which $n$ values the domain wall separates). In the DQC scenario these $\pi/2$ 
domain walls are indeed favored over $\pi$ domain walls \cite{Levin04}, 
as has  been explicitly observed in studies of $J$-$Q$ models \cite{Shao15} (where an 
induced $\pi$ domain wall splits up spontaneously into a pair of $\pi/2$ domain walls), and the domain wall thickness diverges as the DQC point is 
approached from the cVBS side. 

The domain-wall broadening also corresponds to a lowering of an effective potential $V(\phi)=-V_0\cos(4\phi)$ 
experienced by the global cVBS angle in a finite system, which implies that the distribution $P(\phi)$ of the global angle becomes increasingly 
flat as the DQC point is approached. The emergent U(1) symmetry close to the DQC point has been frequently studied by investigating such distributions 
\cite{Sandvik07,Jiang08,Sandvik12}, and the fact that the symmetry has been observed on very large length scale is one of the most important indicators 
of the DQC phenomenon actually being realized in these models \cite{Sachdev08}. 

The emergent U(1) symmetry of the cVBS is analogous to the classical 3D clock models with a potential $-h_q\cos(q\theta_i)$ (with integer $q$) added
to the standard XY model with nearest-neighbor interactions $-J\cos(\theta_i-\theta_j)$ \cite{Levin04}. In the clock models, for $q \ge 4$ the clock term is a 
``dangerously irrelevant'' perturbation, meaning that there is still a phase transition in the standard 3D XY universality class, at some critical 
temperature $T_c$ that depends on $h_q$, but in the ordered state the broken symmetry changes from U(1) to $Z_q$ (i.e., the ordered state corresponds 
to a different renormalization-group fixed point). The emergent U(1) symmetry is associated with a second divergent length scale $\xi'$, which diverges 
faster than the conventional correlation length $\xi$ as $T \to T_c$ from below. The correlation length and the U(1) length are governed by exponents 
$\nu'$ and $\nu$, respectively, and $\nu'>\nu$. The exponents are related to each other and to the scaling dimension of the clock perturbation in a 
way that has been controversial \cite{Oshikawa00,Lou07,Leonard15,Okubo15} and for which new insights were presented very recently \cite{Shao20}. 
The analogous U(1) length scale has also been studied in the standard $J$-$Q$ model \cite{Lou09}, though not yet at the level of precision as was 
possible in the classical case.

\begin{figure}[t]
\includegraphics[width=80mm]{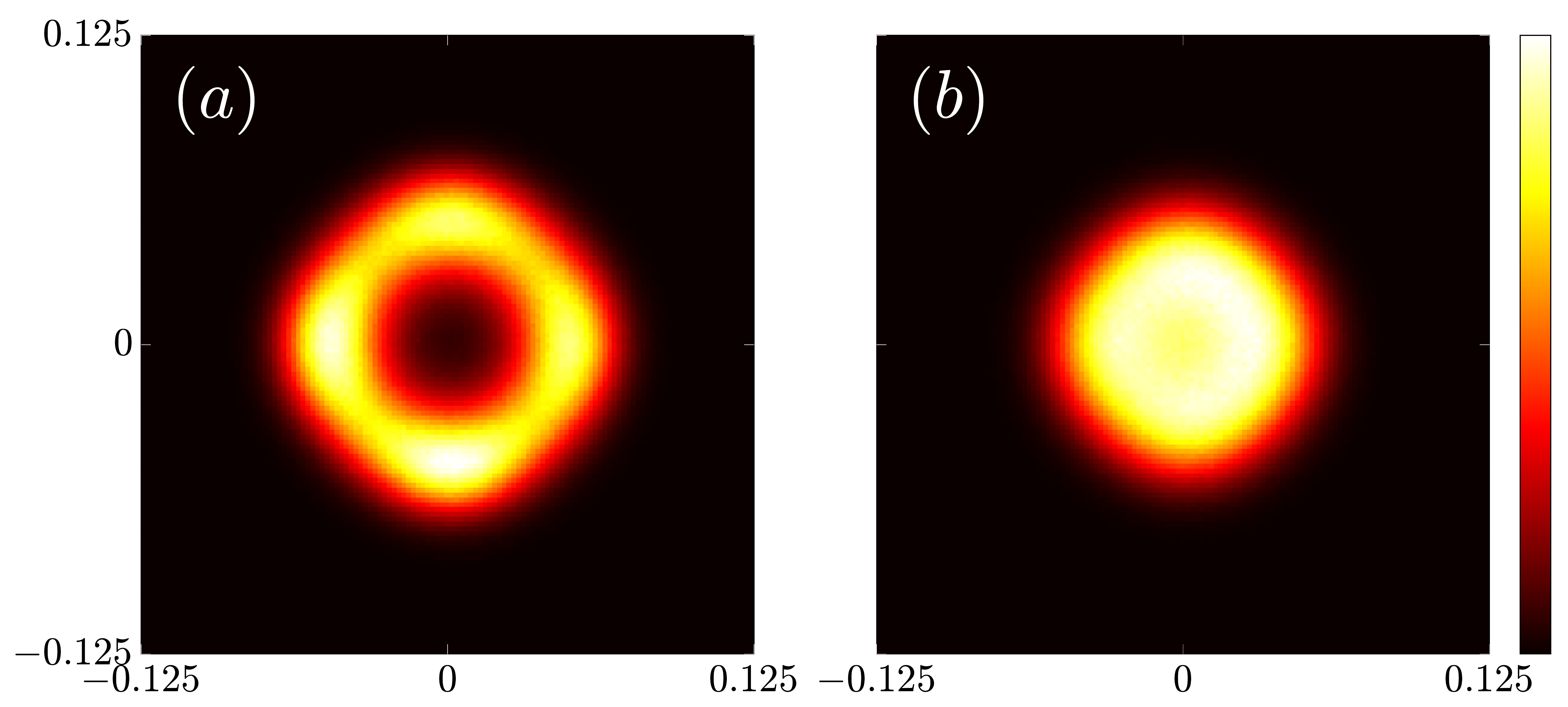}
\caption{Distribution of the cVBS order parameter defined in Eq.~(\ref{mcdef}) for $L=32$ systems with $h=0$ at $g=0.80$ in (a) and $g=0.78$ in (b).
In these calculations $\beta=4L$. The horizontal and vertical axes correspond to ${\rm Re}\{m_c\}$ and ${\rm Im}\{m_c\}$, respectively. The color scale 
corresponds to a linear mapping from $0$ to maximal probability density in each of the panels. The deviations from perfect $Z_4$ symmetry are due to relatively 
short simulations with slow migration of the VBS angle $\phi$.}
\label{fig:z4fluct}
\end{figure}

Here we wish to make some observations regarding the emergent U(1) symmetry of the cVBS as the transition changes from continuous to first-order
as we increase the tuning parameter $h$ (as demonstrated above in Figs.~\ref{Fig:DetailedOrderPara_0} and \ref{Fig:DetailedOrderPara_1}). 
We note that the symmetry broken by the cVBS is normally regarded as $Z_4$, though strictly speaking, from the purely group-theoretical perspective, 
it can also be classified as $Z_2 \times Z_2$, as discussed in detail in Ref.~\cite{Takahashi20}. What qualifies the classification physically as 
$Z_4$ is the nature of the dominant fluctuations between the four degenerate dimer patterns \cite{Levin04}. As discussed above, the elementary domain 
walls are of the $\pi/2$ type, and this means that a state with global angle $n\pi/2$ most easily tunnels to an adjacent state with $(n\pm 1)\pi/2$ (on 
a finite system, where the symmetry is not strictly broken and such fluctuations take place), instead of dominant fluctuations between states with
different $n$ of the type $0 \leftrightarrow 2$ and $1 \leftrightarrow 3$. If the latter type of fluctuations were dominant, the symmetry should physically
be classified as $Z_2 \times Z_2$. 
To be more precise, such fluctuation pattern is equivalent to that when a $Z_2$ symmetry breaks twice, 
which should obviously be regarded as $Z_2 \times Z_2$ symmetry breaking. 

It can be noted that, the classical 3D $q=4$ clock model with potential depth $h_4 \to \infty$ (or, equivalently, the ``hard'' clock model where the angles are 
constrained to the strictly horizontal and vertical directions) in fact maps onto two decoupled Ising models, thus breaking $Z_2 \times Z_2$ 
symmetry in an phase transition of the Ising type. In contrast, the ``soft'' $q=4$ clock model with small $h_4$ exhibits clock-like fluctuations, 
and its phase transition is in the XY universality class on account of the emergent U(1) symmetry. The exact value of $h_4$ at which the change
in fluctuation paths takes place is not known, because difficulties in analyzing the emergent U(1) symmetry when the scaling dimension $y_4$
of $h_4$ is only very slightly negative ($y_4 \approx -0.1$) \cite{Shao20,Pujari15}.

\begin{figure}[t]
\includegraphics[width=80mm]{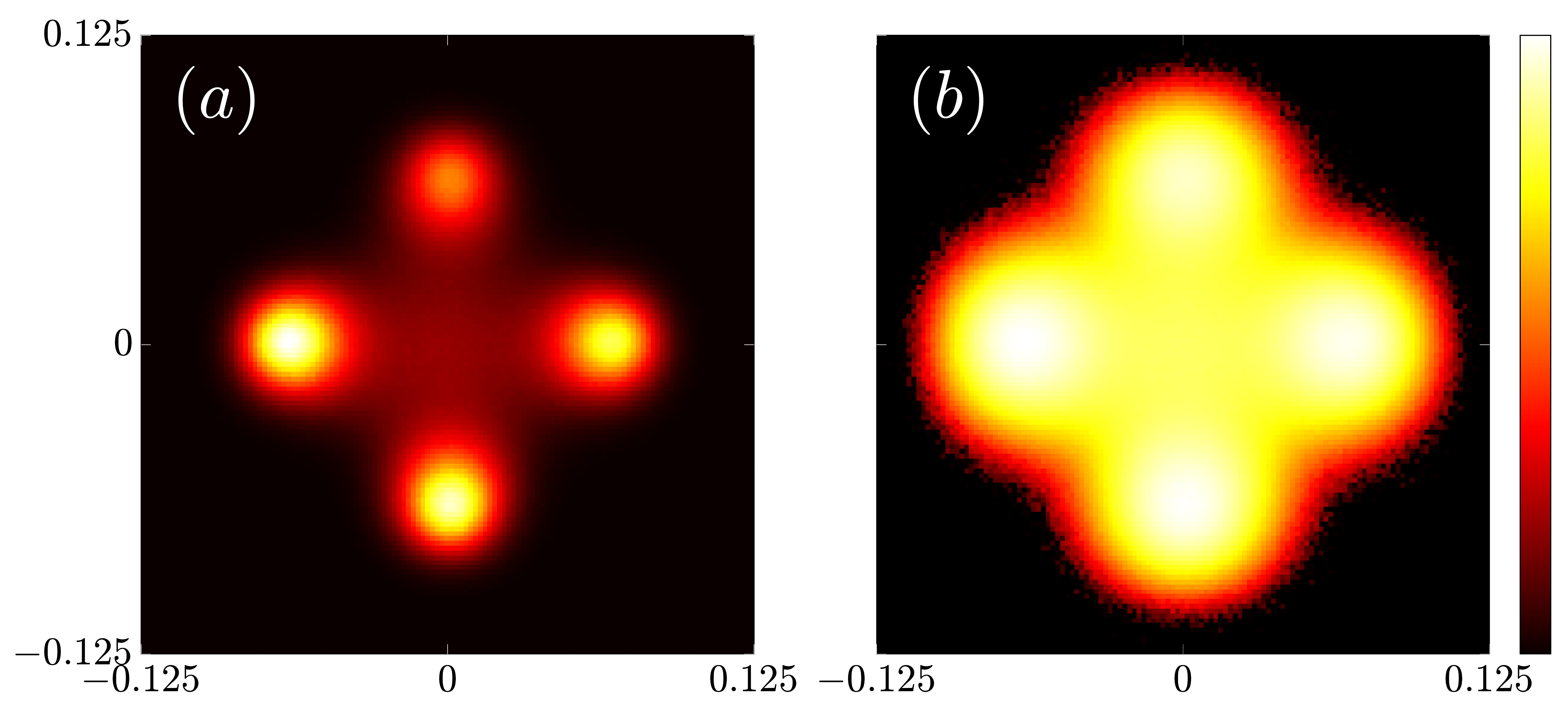}
\caption{Distribution of the cVBS order parameter as in Fig.~\ref{fig:z4fluct}, but for an $L=32$ system with $h=0.2$ at $g=0.68$ and $\beta=4L$. 
In (a) a linear color scale is used to represent the probability density, while in (b) a logarithmic scale is used in order to make better visible 
the density between the four peaks. In (b) the bottom of the color scale (black) represents histogram bins with none or one count among the $10^6$ 
sampled SSE configurations}
\label{fig:z2fluct}
\end{figure}

The cVBS state in the conventional $J$-$Q$ models also exhibit clock-like fluctuations and emergent U(1) symmetry, and is, thus, more similar to 
the soft $q=4$ clock model (though the universality class is different; DQC instead of XY) \cite{Sandvik07,Jiang08,Sandvik12}. In Fig.~\ref{fig:z4fluct} 
we demonstrate these behaviors also for our model at $h=0$, where according to the results presented above in Sec.~\ref{sec:afmcvbs} the AFM--cVBS 
transition appears to be continuous. Fig.~\ref{fig:z4fluct}(a) is for a case where the system is already quite close to the transition into
the AFM state, and for a relatively small system $L=32$, where both the angular and amplitude fluctuations are large. The probability density
is very low in the center of the distribution, reflecting a robust non-zero magnitude $|m_c|$ of the order parameter. The four-fold symmetry
is observed clearly even though all angles $\phi \in [0,2\pi]$ have significant probability density. In Fig.~\ref{fig:z4fluct}(b) the system
is closer to the transition but still slightly inside the cVBS phase, as reflected in the fact that the radial distribution still is maximum
for $|m_c|>0$, though now the density at $|m_c|=0$ is also large. There is still some remnants of $Z_4$ symmetry here (a slightly diamond-like
deviation from circular symmetry of the distribution), which would further diminish if we move even closer to the transition. Proper analysis of the 
size dependence of the distribution \cite{Zhao19,Takahashi20} would be required to draw firm conclusions, but at least it appears plausible here 
that the system has the emergent U(1) symmetry expected at a DQC transition.

It is now interesting to see how the cVBS order parameter distribution evolves as we increase $h$ and enter the regime where the AFM--cVBS
transition is clearly first-order. In Fig.~\ref{fig:z2fluct} we show results at $h=0.2$. Here we no longer observe clock-like fluctuations, 
but instead see significant probability for fluctuations toward the center of the distribution, where in the $Z_4$ case there is very little
weight. If these fluctuations would be due to direct fluctuations between the two maxima on the $x$ or $y$ axis, we would conclude that the
$\pi$ domain wall has replaced the $\pi/2$ one as the softest domain wall, and, therefore, that the fluctuation paths correspond to $Z_2 \times Z_2$ 
symmetry instead of $Z_4$. However, the situation is made more complicated by the fact that coexistence of the AFM and cVBS order parameters
at the first-order transition implies fluctuation paths also into the AFM phase if we are sufficiently close to the phase transition. Indeed, 
upon closer examination of the distribution represented in a higher-dimensional space reveals that the fluctuation paths in Fig.~\ref{fig:z2fluct} 
correspond to significant population of the AFM phase, in which of course the cVBS order parameter is small and weight is produced at the center 
of the distribution. Therefore, from these results we can only conclude that the clock-like $Z_4$ fluctuations are suppressed as the first-order 
cVBS--AFM transition is approached and, therefore, there are no longer any signs of emergent U(1) symmetry. Further inside the cVBS phase, even
for large $h$, we still observe the clock-like $Z_4$ fluctuations.

\subsection{sVBS fluctuations in the cVBS phase}
\label{sec:svbsfluct}

An at first sight puzzling aspect of the Binder cumulant results in Fig.~\ref{Fig:Q4OrderPara} is that the value of $U_s$ is not close to $0$
in the cVBS phase (and also does not approach $0$ for larger system sizes). Normally an appropriately defined cumulant $U_X$ approaches zero 
with increasing system size in a phase where there is no order of the $X$ type, on account of Gaussian fluctuations. However, the sVBS 
fluctuations are not Gaussian-distributed in the cVBS phase, as we show here. 

\begin{figure}[t]
\includegraphics[width=80mm]{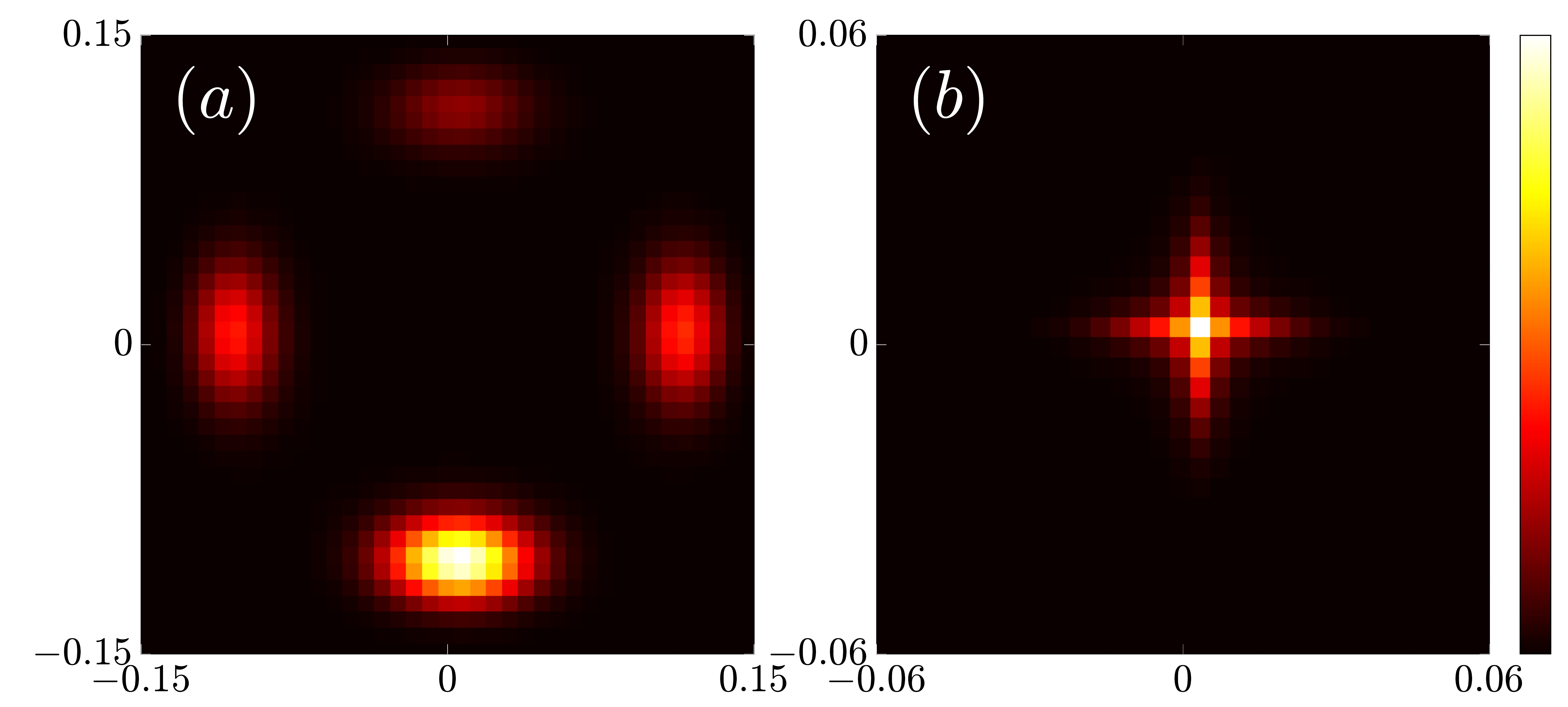}
\caption{Distribution of the cVBS (a) and sVBS (b) order parameters for an $L=16$ ($\beta=4L$) system inside the cVBS phase, at $(g,h)=(0.7,0.4)$. The
distributions combine results of 10 independent runs, all of which were individually stuck in one out of the four cVBS patterns. The
different weight of the four cVBS peaks reflect different numbers of runs stuck in the different patterns.}
\label{fig:svbsfluct}
\end{figure}

Consider the two-dimensional distribution of the complex order parameter $m_s$ defined in Eq.~(\ref{msdef}). Inside the cVBS phase, if the
four-fold symmetry is broken and the system is locked into one of the cVBS pattern, the sVBS order parameter is not isotropic, because
the columnar bond pattern will favor short-range sVBS order with the dimers oriented in the same ($x$ or $y$) direction as that in the cVBS pattern.
Therefore, the conditional sVBS distribution given one of the sVBS pattern will be elongated in one direction, and when averaging
over all the four cVBS pattern a + shaped sVBS distribution is generated. This is demonstrated in Fig.~\ref{fig:svbsfluct} for our model at 
a point rather deep inside the cVBS phase. Since the + shaped distribution is far from a 2D Gaussian, the cumulant will not approach zero 
in this case, thus explaining the results in Fig.~\ref{Fig:Q4OrderPara}. These arguments also apply to cVBS fluctuations inside the sVBS phase, 
and, indeed, in Fig.~\ref{Fig:Q4OrderPara} we can also see that the cVBS cumulant $U_c$ is not zero inside th sVBS phase.

\section{SU(N) model and results for $N=3$}
\label{Sec:SU3Model}

\begin{figure}[t]
\includegraphics[width=70mm,clip]{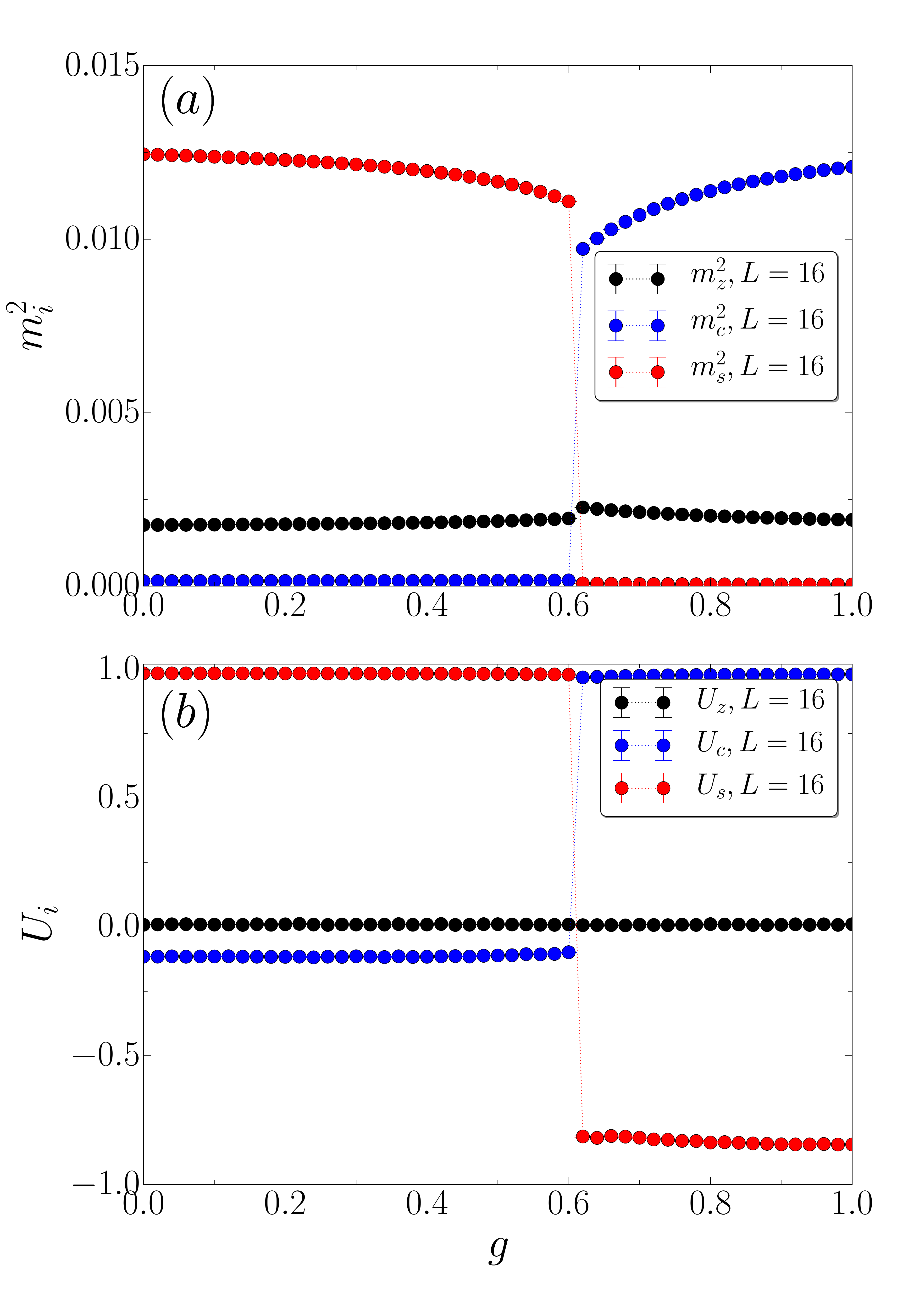}
\vskip-2mm
\caption{Order parameters and corresponding Binder cumulants of the SU(3) variant of the $Q_{3{\mathrm C}}$-$Q_{3{\mathrm S}}$ model, Eq.~(\ref{Eq:Model_Q3Q3s}).
The results were obtained by SSE QMC calculation with $\beta = L$. The outlier point at $g=0.82$ represents a system stuck in a meta-stable state with a
defect in the cVBS order.}
\label{Fig:SU3_OrderPara}
\end{figure}

So far, the model we have studied was defined with the standard $S=1/2$ spin operators with SU(2) symmetry. Spin models on bipartite lattices
have a natural SU(N) generalization, in which the spins on one sublattice transform under the fundamental representation of the group and the ones on the
opposite sublattice transform with the conjugate of the fundamental representation \cite{Read89}. In the Heisenberg model with only nearest-neighbor
Heisenberg exchange, a large-$N$ mean-field-like theory with $1/N$ corrections predicts that AFM order vanishes above $N \approx 4.5$ \cite{Read90}, 
after which the ground state is a cVBS state. QMC simulations for integer $N$ indeed found AFM order up to $N=4$ and cVBS order for larger $N$ \cite{Harada03}. 
QMC simulations with loop-type algorithms can also be generalized to non-integer $N$, and a critical $N$, $N_c \approx 4.5$, in good agreement with 
the $1/N$ expansion was found \cite{Beach09}. $J$-$Q$ models generalized to SU(N) have likewise been useful for testing $1/N$ expansions within 
the DQC scenario \cite{Kaul13},  and a remarkable agreement has been found between the $1/N$ results and QMC finite-size scaling results 
for moderate values of $N$ \cite{Kaul12,Dyer15}.

Given that larger $N$ drives the Heisenberg and $J$-$Q$ systems toward VBS ordering, in the model studied here we would expect the
AFM phase to shrink upon increasing $N$. Here our aim is just to demonstrate this behavior, and we leave more detailed studies to future work.
Using $N=3$, we set $h=0$, i.e., we only have the competing $Q_{3{\mathrm C}}$ and $Q_{3{\mathrm S}}$ left, as in the Hamiltonian Eq.~(\ref{Eq:Model_Q3Q3s}). 

As shown in Fig.~\ref{Fig:SU3_OrderPara}, the SU(3) system undergoes a clearly first-order direct sVBS--cVBS transition. Here the VBS order 
parameters are defined based on the natural generalization of the singlet projection operator and the AFM order parameter is defined by the imbalance
of the different ``color" on each site \cite{Kaul12}. We use the same definitions of the Binder cumulants as in Eq.~(\ref{udefs}). The cumulants $U_c$
and $U_s$ will approach $1$ if there is cVBS and sVBS long-range order, respectively. If there is no such order the values will not go to $0$, however,
for the same reasons as discussed in Sec.~\ref{sec:svbsfluct}. In the case of AFM order, the definition in Eq.~(\ref{uzdef}) does not account properly
for SU(3) symmetry and $U_z$ will approach a value different from $1$. In a non-AFM state the value should still approach $0$. The results shown in 
Fig.~\ref{Fig:SU3_OrderPara} demonstrate a strongly first-order, direct transition between the two VBS states. The AFM order parameter is small and 
vanishes upon increasing the system size, and the value of $U_z$ stays very close to $0$.

We can conclude that the AFM order is always suppressed in the SU(3) $Q_{3{\mathrm C}}$-$Q_{3{\mathrm S}}$ model and there is a first-order sVBS--cVBS transition
similar to the case of our SU(2) model in Fig.~\ref{Fig:PhaseDiagram} when $h$ is large. We can then achieve a similar phase diagram with an AFM phase and a triple
point by adding Heisenberg exchange $J$ to the SU(3) Hamiltonian. As mentioned above, for $N \ge 5$ the generalized Heisenberg model is always cVBS ordered, and previously
a second-neighbor {\it ferromagnetic} Heisenberg term (i.e., on the diagonals of the $2\times 2$ plaquettes on the square lattice) 
was added in order 
to enhance AFM order and realize an AFM--cVBS transition for any $N \ge 5$ (and for any $N \ge 2$ if also a $Q$ term is used) \cite{Kaul12}. In our model 
studied here, the $Q_{3{\mathrm S}}$ interactions compete agains cVBS order and therefore the SU(N) $J$-$Q_{3{\mathrm C}}$-$Q_{3{\mathrm S}}$ model should have a phase diagram similar 
to Fig.~\ref{Fig:PhaseDiagram} (with $h$ replaced by $J$ for $N>2$). Most likely, we should also then be able to tune the AFM--cVBS from a continuous 
DQC transition to a first-order transition for any $N$.

\section{Summary and Discussion}
\label{Sec:Discussion}

In this paper we have proposed a sign-free designer quantum spin model, Eq.~(\ref{Eq:Model_Q4}), that enables large-scale QMC studies of competing VBS
states and different tupes of AFM--VBS transitions. In this section, we first summarize our main results and conclusions in Sec.~\ref{Subsec:Summary}, and
then in Sec.~\ref{Subsec:Fractons}, inspired by the proposal in Ref.~\onlinecite{You20} and related arguments in Ref.~\onlinecite{Takahashi20}, we further
discuss possible fracton properties of the spinons in the sVBS phase. Finally, we will discuss future prospects in Sec. \ref{Subsec:Future}. 
              		
\subsection{Conclusions}
\label{Subsec:Summary}

Our QMC results show that the $Q_{\rm 3C}$-$Q_{\rm 3S}$-$Q_{\rm 4H}$ model has a ground state phase diagram with several interesting features. 
While the $Q_{\rm 3C}$ and $Q_{\rm 3S}$ interactions compete, inducing columnar and staggered VBS correlations, respectively, the $Q_{\rm 4H}$ term is
favorable to both kinds of orders but suppresses AFM ordering. The AFM phase separating the cVBS and sVBS phases when $Q_{\rm 4H}$ is absent or small
therefore shrinks as $Q_{\rm 4H}$ is increased, and eventually the model undergoes a direct phase transition between the cVBS and sVBS phases. 
The sVBS--cVBS and sVBS--AFM transitions are both first-order, as expected from standard Landau arguments. In contrast, the AFM--cVBS transition appears
to be a Landau-forbidden continuous one for small $Q_{\rm 4H}$, i.e., when the sVBS phase is far away, while it becomes clearly first-order when $Q_{\rm 4H}$ is
increased and the two VBS phases approach each other. Thus, there is a triple point where the three first-order phase boundaries meet each other, and most
likely there is a tricritical point on the AFM--cVBS boundary. The continuous AFM--cVBS transitions should be of the DQC type, and the existence of a tricritical
point would imply that the nature of the transition can be drastically changed without changing the symmetry of the model, i.e., that symmetry-allowed
terms can drive the system away from the DQC fixed point.

Recently, the so-called ``walking'' behavior in non-unitary complex CFTs \cite{Wang17,Gorbenko18} has gained attention as a way to explain the weakly
first-order nature of some transitions, such as $q$-state Potts models with $q>4$. This scenario has also been invoked by proponents of a generic weakly
first-order AFM--VBS transitions \cite{Wang17,Ma19b,Nahum19}.  Our model is the first in which the AFM--cVBS transition can be continuously tuned
from what appears to be continuous to clearly first-order. This model should therefore constitute a good framework for studying the walking behavior in detail. 
The initial results presented here show that the emergent U(1) symmetry of the order parameter is absent when the transition becomes first-order. This
behavior is in sharp contrast to the emergent higher symmetries, O(4) and SO(5), of the order parameters recently observed at some and first-order
AFM--VBS transitions in related models \cite{Zhao19,Serna19,Takahashi20}.

We also discussed a generalized model with SU(N) symmetry and showed some results for the SU(3) case. It is well known that AFM order is suppressed
when $N$ is increased \cite{Read89,Read90,Harada03}, and in the SU(3) $Q_{\rm 3C}$-$Q_{\rm 3S}$ model the AFM phase existing in the SU(2) case is already
absent. AFM order can be brought back by including a Heisenberg ($J$) term (possibly with longer-range interactions for larger $N$ \cite{Kaul12}). Thus, we
can in principle study phase diagrams and investigate the evolution of different phase transitions similar to what we did here for the SU(2) model.
               
\subsection{Fractons}
\label{Subsec:Fractons} 

\begin{figure}[t]
\includegraphics[width=63mm,clip]{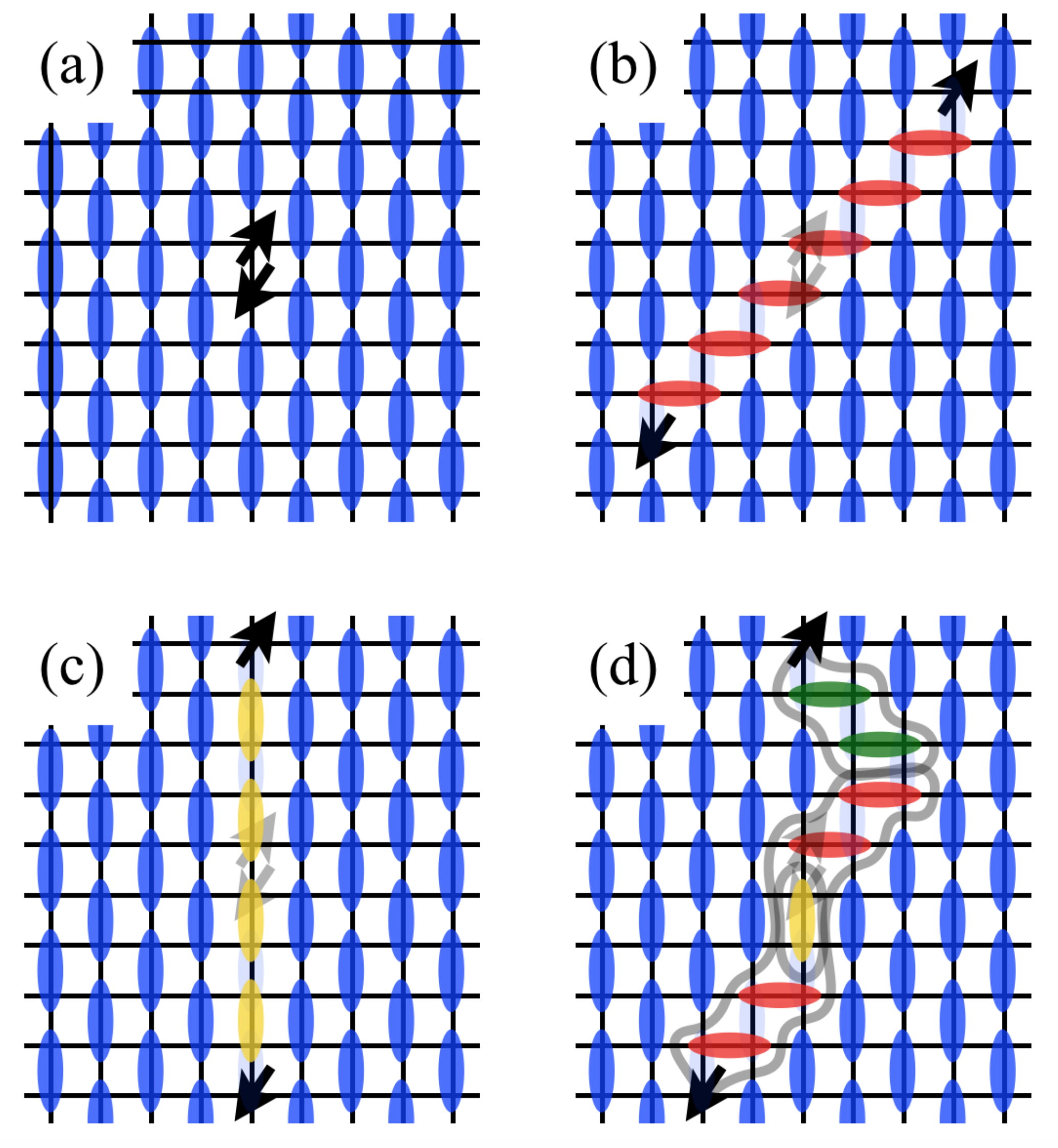}
\caption{Schematic depiction of triplon excitations in the sVBS phase. 
(a) shows a simple triplon excitation where a single valence bond is broken, and (b,c) show examples of spinons propagating in
different directions. In (d) the spinons move by a combination of moves from (b) and (c). The original position of the spinons and the
original valence bond patterns are shown with light colors. Domain walls are indicated by gray lines in (d).}
\label{Fig:sVBSspinons}
\end{figure}

Recently, there have been discussions on the mobility properties of spinons in a 2D VBS state different from the ones considered here---the pVBS phase.
Within a scenario in which the plaquette singlets are treated as rigid objects, spinons as well as their bound ``triplon'' states are fractons,
restricted to motion in one dimension \cite{You20}. One prediction following from this picture is that spinons do not deconfine, and instead of a
DQC transition (which in the DQC theory can take place in a cVBS or a pVBS) a first-order transition results from nucleation of triplons.
However, the fracton argument for pVBS should only apply in extreme cases, as in generic quantum spin models the four-spin singlets can not
be regarded as rigid objects but must be allowed to decay into dimer singlets \cite{Takahashi20}. Here we note that some of the prerequisites
for fractons outlined in Ref.~\cite{You20} in the context of the pVBS has analogies in the sVBS, and the latter may be a more likely host of
fractons under realistic conditions. Though we do not have quantitative results from QMC simulations, we will here argue that fracton-like
excitation are generically possible in sVBS states.

The fully mobile spinons play a crucial role as vortices in the DQC theory, and it is therefore interesting to compare the nature and mobility of the spinons
in cVBS and sVBS states. While the two VBS phases are both four-fold degenerate, they correspond to different kinds of symmetry breaking of the lattice,
thus having different minima in the Landau free energy space. The two VBS phases also host different types of domain walls, leading to different behaviors of
the vortex excitations (Fig.~\ref{Fig:Vertex}). It has been explained before that the geometric structure of the cVBS pattern forces the vortex to always have
a spinon in its core, and this is a crucial aspect of the continuous cVBS--AFM transition of the DQC type \cite{Levin04}. In contrast, the sVBS phase does
not necessarily require spinons-type vortices, though vortices with spinons are also possible (Fig.~\ref{Fig:Vertex}). In the latter case, which to our
knowledge has not been previously considered in this context, these spinons may be regarded as fractons, similar to the spinons recently discussed in the pVBS
phase \cite{You20}.

\begin{figure}[t]
\includegraphics[width=68mm,clip]{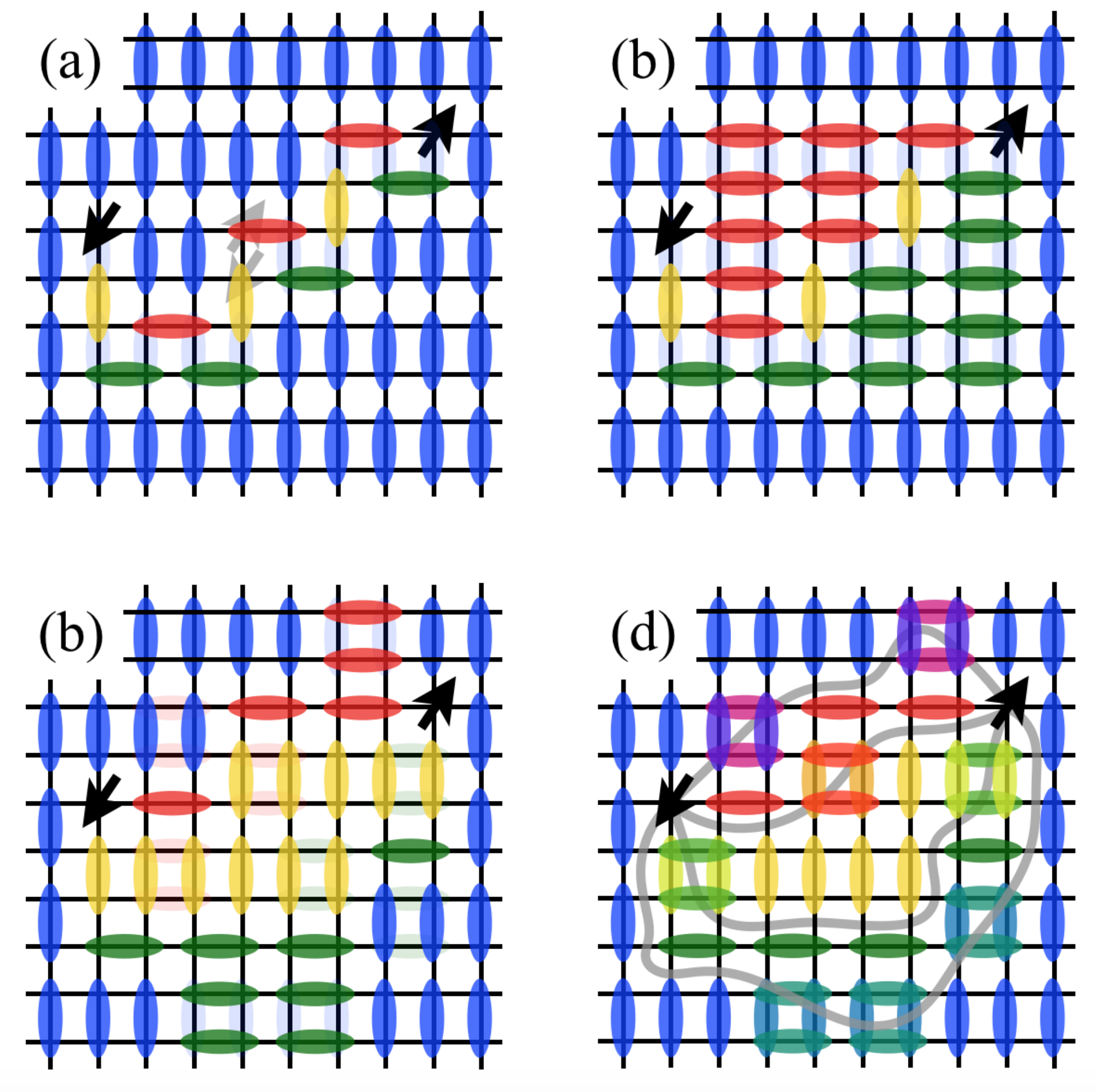}
\caption{Depiction of triplon excitations in the cVBS phase. Panel (a) shows an example of a combination of moves induced by $J$-terms, in analogy
with Fig.~\ref{Fig:sVBSspinons}(d) in the sVBS phase. Parts (b) and (c) show how local fluctuations can alter the configuration shown in (a)
without changing the position of the spinons.  The original position of the spinons and the valence bond patterns are shown with light colors
in (a-c).  In (d) the $\pi/2$-domain walls are indicated with gray curves.}
\label{Fig:cVBSspinons}
\end{figure}

Figure \ref{Fig:sVBSspinons} shows some possible triplon excitations in the sVBS phase. Fig.~\ref{Fig:sVBSspinons}(a) shows a single valence bond
in the sVBS phase being broken, creating a gapped triplon excitation. By applying $J$-terms in the Hamiltonian, we can see that the individual spinons
can move around in two different directions; along the 45$^{\circ}$ diagonal in Fig.~\ref{Fig:sVBSspinons}(b), or vertically along the direction of original
sVBS dimers in Fig.~\ref{Fig:sVBSspinons}(c). In Fig.~\ref{Fig:sVBSspinons}(d) we show how the spinons can move around by combinations of those two
different moves. However, such combinations of moves in different lattice directions result in multiple domain walls depicted as gray lines in 
Fig.~\ref{Fig:sVBSspinons}(d). There are two different kinds of possible domain walls, $\pi/2$-domain walls where a vertical sVBS pattern is adjacent to a 
horizontal sVBS pattern (corresponding to $\pi/2$ rotation in the order parameter space), or $\pi$-domain walls otherwise. In the diagrams, we indicate
the $\pi/2$-domain walls with gray lines, and two such lines amounts to a $\pi$-domain wall.

At an AFM--cVBS DQC point, it is predicted \cite{Senthil04} and numerically observed \cite{Shao15} that the domain wall energy vanishes, resulting in
configurations such as those depicted in Fig. \ref{Fig:cVBSspinons} to contribute substantially to low-lying excited states. The spinons can then move around
essentially freely when the cVBS order weakens, hence the DQC phenomenon.
In the case of the sVBS, we can see in Fig.~\ref{Fig:sVBSspinons}(d) that there would be two different
types of domain walls, which must in general have different domain wall energies. Therefore, only either (b) $\pi/2$- or (c) $\pi$-domain walls would
contribute to the low-lying state, and configurations like (d) would be suppressed because of the presence also of the non-favored type of domain walls
(and which one has a lower energy should depend on the specific microscopic interactions). 
As a consequence the spinons are only able to move freely in one or two directions, depending on the favored domain wall type, becoming effectively fractons. 

The fracton property is emergent, and in principle, the spinons can still move around in different directions in the case where the $\pi/2$-domain wall
in Fig.~\ref{Fig:sVBSspinons}(b)
is preferred. For example, the spinons can not only move apart, but also come back together at a location which is not necessarily the original location
where the excitation was originally created. Subsequently, the spinons can move in the {\it other} diagonal direction, and again result in a different
position as a pair triplon.  Overall, the triplon can move around in all directions, but the most dominant mobility of the individual spinons should be
mainly restricted to the one-dimensional sublattice, and this constraint must suppress the emergent U(1) symmetry and have an affect on the nature of the
phase transition. As argued in Ref.~\cite{You20} in the case of the pVBS, one can expect a first-order transition due to nucleation of spinons. The
details of the nucleation process should be different in an sVBS state, as we discuss next.

\begin{figure}[t]
\includegraphics[width=65mm,clip]{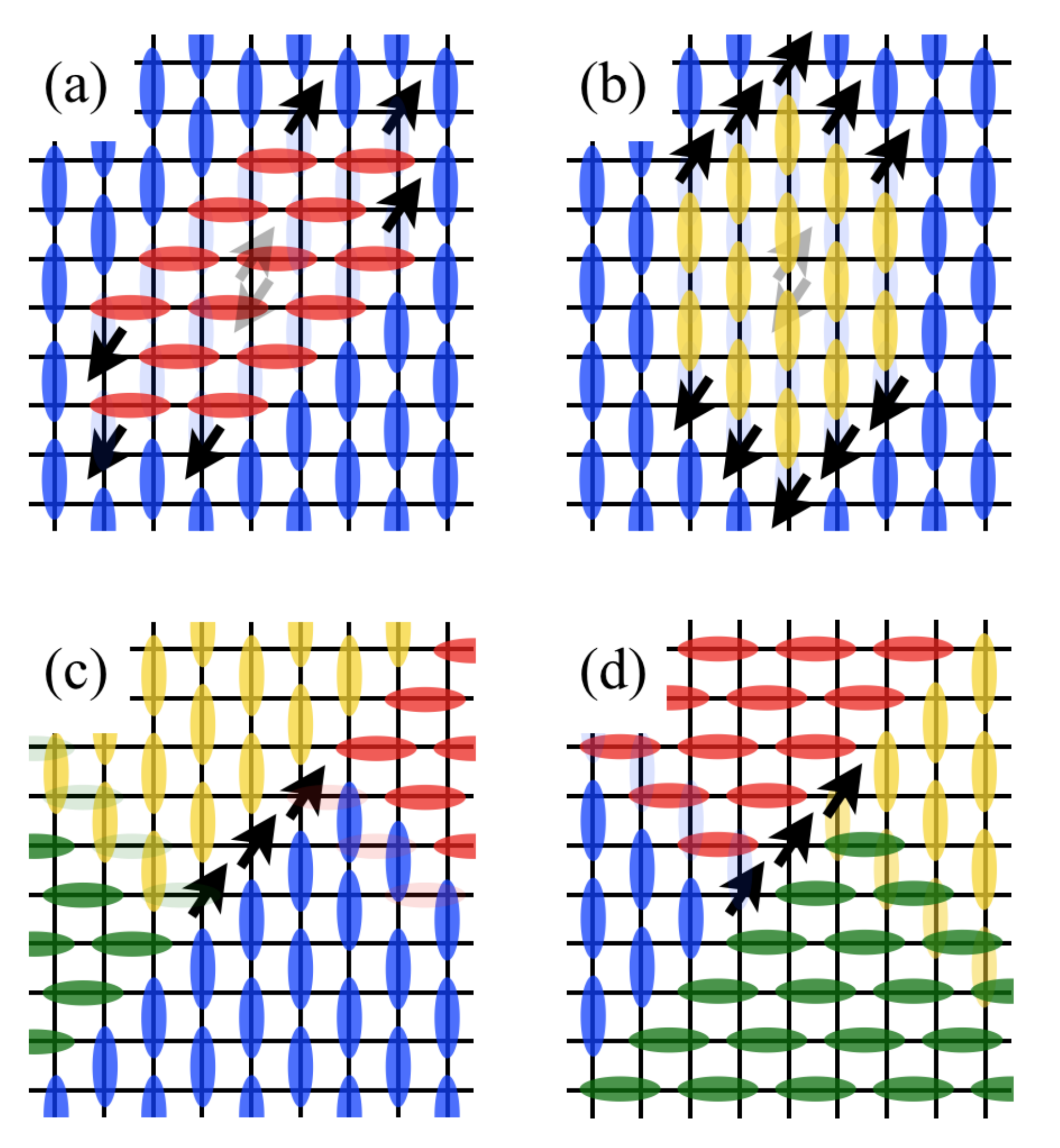}
\caption{Depiction of excitations and defects with multiple spinons in the sVBS phase. Configurations (a) and (b) both illustrate
how thickening domains initially caused by two spinons in Fig.~\ref{Fig:sVBSspinons}(b,c) must induce multiple spinons at both edges.
The configurations in  (c) and (d) show that vortices can also in principle host multiple spinons in their cores.}
\label{Fig:MultiSpinons}
\end{figure}

The mobility properties of the spinons in the sVBS phase we have discussed now is in sharp contrast with that in the cVBS phase. As we show in
Fig.~\ref{Fig:cVBSspinons}, in the well understood cVBS phase the resulting configuration can easily fluctuate locally by applying the $J$-terms. 
This mechanism only requires $\pi/2$-domain walls, which is indeed the one that is favored energetically. We can also see from Fig.~\ref{Fig:cVBSspinons}
that the local fluctuation also allows the domain walls to thicken in the cVBS phase, thus creating alternating domains with finite area. This again
contrasts with sVBS, because, as we show in Figs.~\ref{Fig:MultiSpinons}(a,b), thickening domain walls into extended domains will necessarily require
the creation of more spinons. The staggered pattern in the phase also allows the vortices to be occupied by multiple spinons, as shown in
Figs.~\ref{Fig:MultiSpinons}(c,d), which again is in sharp contrast to the cVBS case. 
Although in the general case it is likely that spinless vortices would be energetically favored \cite{Banerjee11}, 
this property of allowing multiple spinons may play a role for the first-order transitions taking place,
since it allows nucleation of AFM order from clusters of spinons. 

The true lowest triplon excitation state should be a superposition of the type of extended triplet valence-bond basis states discussed above, 
each of them transforming to a different constituent of the state when a Heisenberg $J$-term is applied. The specific model discussed in the
preceding sections does not have any explicit $J$ terms (two-spin singlet projector) but the same fluctuation effect is also achieved by appropriately
applying $Q$ terms. Valence-bond configurations with higher diagonal energy should have a smaller contribution to the lowest triplon excitation, 
and by analyzing those configurations we can argue how the spinons proliferate in space. Though we have not discussed the mechanisms in a rigorous
manner here, the picture is intuitive and should capture the correct physics qualitatively.
		
\subsection{Future prospects}
\label{Subsec:Future}

The work we presented here illustrates the power of the designer Hamiltonian approach in engineering sign-free Hamiltonians exhibiting interesting
phase transitions and enabling unbiased numerical studies of physically interesting situations. Our results suggest several possible follow-up studies,
some of which we summarize here.
		
First, precisely determining the tricritical point in Fig.~\ref{Fig:PhaseDiagram} where the cVBS--AFM transition changes from continuous to first order is an
important task. If the emergent U(1) symmetry also disappears at the multicritical point, this would indicate a compelling correspondence between the emergent
U(1) and continuous DQC transitions. Also, calculating the critical exponents along the critical line and comparing it with existing results for the
conventional $J$-$Q$ model \cite{Shao16} would be helpful to confirm the universality class of DQC point. Here we should again note that the continuous
transition in the conventional $J$-$Q$ model has not been demonstrated completely conclusively, though there are no explicit signs of first-order
discontinuities. Detailed studies of the putative tricritical point, or, alternatively, demonstrating its absence (i.e., a wek first-order transition
below the blue circle indicating the putative tricritical point in Fig.~\ref{Fig:PhaseDiagram}) would also be very useful in this regard.
		
We have discussed the fluctuation patterns of the order parameter inside the VBS phases. The change in fluctuation paths where the transition
changes to first order can be seen as a numerical confirmation of the graph-theoretic approach introduced in the context of the AFM--pVBS
transitions in Ref.~\onlinecite{Takahashi20}. Further studying when and how the fluctuation pattern between degenerate ground states
changes within a single phase could shed light on what kind of phases are allowed to exist adjacent to each other in the phase diagram,
and indicate relevant and irrelevant perturbations of the DQC point.
		
Another interesting aspect of the sVBS phase deserving further study is the nature of the spinons and their bound states. As we show in
Fig.~\ref{Fig:MultiSpinons}, the vortices in the sVBS phase can host one or several spinons, as well as no spinon. It would be very interesting
to test in numerical simulations whether the multi-spinon configurations actually appear in the vicinity of the cVBS--sVBS or AFM--sVBS phase
transitions, and to investigate how well the fracton picture explains the behaviors we have observed here. It should already be clear from the
arguments regarding the triplon excitations shown in Fig.~\ref{Fig:sVBSspinons} that the spinon mobility is very anisotropic, and studying dynamical
aspects of the sVBS phase near criticality, e.g., the dynamic spin and dimer structure factors (using methods discussed in
Refs.~\onlinecite{Qin17b,Shao17}) should shed further light on the nature of the excitations.

%		Finally, the fact that a well-designed term in the $J$-$Q$ Hamiltonian framework 
%		can induce direct cVBS--sVBS phase transition, opens up new possibilities 
%		of studying a lot of previously unreachable direct phase transitions. 
%		For example, a model that will favor an sVBS vortex with a single spinon, may have 
%		similar AFM--sVBS transition properties as DQC, since now the topological defect in the sVBS pahse
%		possesses spinons carrying AFM order as they do in the cVBS phase. 
%		Designing such models with exotic phase transitions would be an interesting direction to explore.

\begin{acknowledgments}
                We would like to thank Ribhu Kaul for useful discussions.
  		This work was supported by the NSF under Grant No.~DMR-1710170 and by a Simons Investigator Grant.
  		The numerical calculations were carried out on the Shared Computing Cluster managed by Boston University's 
                Research Computing Services.
\end{acknowledgments}

\end{document}